\documentclass[10pt, conference, compsocconf]{IEEEtran}

\usepackage{amssymb,amsfonts}

\usepackage[mathscr]{eucal}

\usepackage{stmaryrd}

\usepackage{amsbsy}

\usepackage{bm}

\usepackage{braket}

\newcommand{\V}[2][]{{\bm{#1\mathbf{\MakeLowercase{#2}}}}} 
 
\newcommand{\Vn}[3][]{{\bm{#1\mathbf{\MakeLowercase{#2}}}}^{(#3)}} 
 
\newcommand{\M}[2][]{{\bm{#1\mathbf{\MakeUppercase{#2}}}}} 
\newcommand{\Mn}[3][]{{\bm{#1\mathbf{\MakeUppercase{#2}}}}^{(#3)}} 
 
\newcommand{\MnC}[4][]{\Vn[#1]{#2}{#3}_{#4}} 
 
\newcommand{\Oprod}{\circ} 
 
\usepackage{cite}

\ifCLASSINFOpdf
\else
\fi
\usepackage[cmex10]{amsmath}
\usepackage[caption=false,font=footnotesize]{subfig}
\usepackage{url}

\usepackage{algpseudocode}

\usepackage{amssymb,latexsym,graphicx,multicol}
\usepackage{setspace}
\usepackage[usenames]{xcolor}
\usepackage{mathtools}
\usepackage{hyperref}
\usepackage{cleveref}
\usepackage{algorithm,algpseudocode}
\usepackage{mathdots}

\usepackage{tikz}
\usetikzlibrary{3d}
\usetikzlibrary{patterns}
\usetikzlibrary{calc}
\usetikzlibrary{arrows}

\usepackage{pgfplots}
\usepackage{pgfplotstable}
\usepackage{etoolbox}
\graphicspath{{fig/}}

\algdef{SE}[FORDOTS]{ForDots}{EndForDots}{$\ddots$}{$\iddots$}

\newcommand{\comment}[1]{}

\newtheorem{theorem}{Theorem}[section]
\newtheorem{corollary}{Corollary}[section]
\newtheorem{lemma}{Lemma}[section]
\newtheorem{definition}{Definition}[section]

\newtheorem{fact}{Fact}[section]

\crefname{theorem}{Theorem}{Theorems}
\crefname{corollary}{Corollary}{Corollaries}
\crefname{lemma}{Lemma}{Lemmas}
\crefname{definition}{Definition}{Definitions}
\crefname{proposition}{Proposition}{Propositions}
\crefname{fact}{Fact}{Facts}
\crefname{algorithm}{Algorithm}{Algorithms}
\crefname{section}{Section}{Sections}
\crefname{line}{Line}{Lines}
\crefname{algorithm}{Algorithm}{Algorithms}
\crefname{equation}{Equation}{Equations}

\newcommand*{\integers}{\mathbb{Z}}

\newcommand{\from}{\colon}

\newcommand{\PP}{\mathcal{P}}

\usepackage{enumerate}

\newcommand{\X}{\mathcal{X}}

\newcommand{\I}{\mathcal{I}}

\newcommand{\mults}{$N$-ary multiplies}

\DeclareMathOperator{\nnz}{nnz}

\begin{document}
%
\title{Communication Lower Bounds for Matricized Tensor Times Khatri-Rao Product}


\author{\IEEEauthorblockN{Grey Ballard and Kathryn Rouse}
\IEEEauthorblockA{Department of Computer Science\\
Wake Forest University\\
Winston Salem NC, USA\\
Email: \{ballard,rousekm\}@wfu.edu}
\and
\IEEEauthorblockN{Nicholas Knight}
\IEEEauthorblockA{Courant Institute of Mathematical Sciences\\
New York University\\
New York NY, USA\\
Email: nknight@nyu.edu}
}


%


\maketitle

\begin{abstract}
The matricized-tensor times Khatri-Rao product (MTTKRP) computation is the typical bottleneck in algorithms for computing a CP decomposition of a tensor. 
In order to develop high performance sequential and parallel algorithms, we establish communication lower bounds that identify how much data movement is required for this computation in the case of dense tensors.
We also present sequential and parallel algorithms that attain the lower bounds and are therefore communication optimal.
In particular, we show that the structure of the computation allows for less communication than the straightforward approach of casting the computation as a matrix multiplication operation.
\end{abstract}


%
\IEEEpeerreviewmaketitle

\section{Introduction}

Tensor decompositions are a powerful tool in the analysis of multidimensional datasets arising from a wide variety of applications.
Two of the most popular decompositions, known as CP and Tucker, are generalizations of the matrix singular value decomposition (or principle component analysis) and form low-rank approximations of tensor data.
They are used heavily in the scientific computing, signal processing, and machine learning communities \cite{KB09,AGHKT14,SL+17}, and the formulations and fundamental algorithms for computing these decompositions are well established.

However, their growing popularity, along with the continued increase in the size of datasets across applications, has increased demand for high-performance parallel algorithms and implementations.
To deliver efficient solutions for tensor problems, high performance computing can leverage the wealth of knowledge and experience with dense and sparse matrix computations, which are closely related to the computational kernels within tensor decomposition algorithms.
In particular, obtaining high performance requires minimizing the cost of data movement among processors and within the memory hierarchy, as the costs of communication are an increasing bottleneck on today's architectures.

The goal of this work is to focus on the communication costs of the bottleneck computation within algorithms that compute the CP decomposition.
The CP decomposition, as we discuss in \cref{sec:prelim}, approximates a tensor as a sum of rank-one tensors, typically represented as a set of factor matrices, much like a low-rank approximation of a matrix.
Nearly all optimization schemes for computing a CP decomposition spend most of their time in a computation known as matricized-tensor times Khatri-Rao product (MTTKRP), and in this work we focus on MTTKRP in the case of dense tensors.
Our results are based on a sequential two-level memory model and a distributed-memory parallel model.

The main contributions of this paper are to 
\begin{itemize}
	\item establish sequential and parallel communication lower bounds for dense MTTKRP (\cref{sec:LBs});
	\item present communication-optimal sequential and parallel dense MTTKRP algorithms (\cref{sec:algs}); 
	\item expose the opportunities within tensor computations to achieve better locality than is available within matrix computations (\cref{sec:disc}).
\end{itemize}
We discuss related work in \cref{sec:related} and conclude the paper in \cref{sec:conc}.

\section{Preliminaries}
\label{sec:prelim}

\subsection{CP Decomposition}
\label{sec:CP}

The CANDECOMP/PARAFAC or canonical polyadic (CP) decomposition is the approximation of a tensor by a sum of rank-one tensors.
Given an $N$-way tensor $\X$ of dimensions $I_1\times \cdots\times I_N$, a rank-$R$ CP decomposition, represented by $N$ factor matrices $\{\Mn{A}{k}\}_{k\in[N]}$, is given by
\[ \X \approx \sum_{r\in [R]} \MnC{A}{1}{r} \Oprod \cdots \Oprod \MnC{A}{N}{r},  \]
where $\MnC{A}{k}{r}$ is the $r$-th column of matrix $\Mn{A}{k}$, or equivalently,
\begin{equation}
\label{eq:CP}
\X(\V{i}) \approx \sum_{r\in [R]} \Mn{A}{1}(i_1,r) \cdots \Mn{A}{N}(i_N,r)\text,
\end{equation}
where $\V{i} = (i_1,\dots,i_N)$.

Computing a CP decomposition involves solving a nonlinear optimization problem to minimize the approximation error, typically measured in the $\ell_2$-norm.
The most common optimization algorithms either use an alternating least squares (ALS) approach or a gradient-based algorithm.
The ALS algorithm alternates among the factor matrices, improving one factor matrix at a time.
When all but one factor matrix are fixed, optimizing the variable factor matrix is a linear optimization problem that can solved in closed form via the normal equations.
In a gradient-based algorithm, the gradients with respect to all factor matrices are computed and used to determine the variable updates.
In both cases, setting up the normal equations and computing the gradient are bottlenecked by a particular computation that involves the tensor and all but one of the factor matrices.
This computation is known as \emph{MTTKRP}. 

\subsection{MTTKRP}
\label{sec:MTTKRP}

MTTKRP inputs an $N$-way tensor $\X$, $N\ge 2$, of dimensions $I_1\times \cdots\times I_N$, a fixed mode $n \in [N]$, and an ($N{-}1$)-tuple of matrices $\{\Mn{A}{k}\}_{k \in [N] \setminus \{n\}}$ each of dimensions $I_k\times R$.
MTTKRP outputs a single matrix $\Mn{B}{n}$, of dimensions $I_n\times R$.
(For a fixed $n$, the matrix $\Mn{A}{n}$ and the superscript on $\Mn{B}{n}$ are irrelevant.)
Throughout the discussion, the underlying set of values is any nonempty set closed under two binary operations, denoted by addition and multiplication, say, the real numbers.
\begin{definition}
\label{def:MTTKRP}
An \emph{MTTKRP algorithm} maps
\[ \Big(\X,\ \{\Mn{A}{k}\}_{k \in [N] \setminus \{n\}}\Big) \mapsto \Mn{B}{n}\text, \]
where for each $(i_n,r) \in [I_n] \times [R]$,
\begin{equation}
\label{eq:MTTKRP}
\Mn{B}{n}(i_n,r) = \sum_{\V{i}} \X(\V{i}) \prod_{k \in [N] \setminus\{n\}} \Mn{A}{k}(i_k,r)\text,
\end{equation}
where summation is over all $\V{i}$ with $n$-th entry $i_n$ in the set
\[ \I = [I_1] \times \cdots \times [I_N] \times [R]\text. \]
The products are evaluated atomically, as \emph{\mults}.
\end{definition}

The atomicity of the \mults\ precludes reusing factors across products,
Moreover, the generality of the arithmetic model precludes a number of other practical optimizations.
For example, assuming existence of a zero element, many operations could be avoided if $\X$ were sparse.
Or, assuming distributivity, the operation count decreases when factoring products through the sums.
Or, assuming the ring axioms hold, Strassen's algorithm could be used in place of the classical matrix multiplication algorithm.
Our ongoing work addresses these optimizations, which change the algorithmic structure in ways that aren't captured by the present lower bound proof approach.

\subsection{Computation Models}

\paragraph{Sequential Model}

Our model sequential machine includes a single processor, connected to two storage devices called \emph{fast} and \emph{slow memory}.
Fast memory can hold up to $M$ values at once, while slow memory has unbounded capacity.
The processor performs (binary) adds and \mults\ on values in fast memory and communicates values between the two memories. 
\emph{Communication} consists of \emph{loads} and \emph{stores}, instructions that read individual values from slow memory and write them to fast memory, or vice versa. 
This model is known as the two-level sequential memory model \cite{BCDH+14} or the I/O complexity model \cite{HK81}.

\paragraph{Parallel Model}

Our parallel model includes $P$ processors, each connected to its own \emph{local memory} and to all other processors via a network.
Local memory holds up to $M$ values, so overall the machine holds at most $PM$ values. 
As in the sequential case, each processor can operate on values in its local memory, while \emph{communication} now consists of \emph{sends} and \emph{receives}, instructions that read individual values from local memory and write them to the network, or vice versa.
We assume each processor can send or receive only one value at a time, but two disjoint pairs of processors can communicate simultaneously.
This model is known as the MPI model \cite{TRG05}, or $\alpha$-$\beta$-$\gamma$ model \cite{BCDH+14}.
In this work, we focus on the amount of data communicated (bandwidth cost) and ignore the number of messages communicated (latency cost). 

\section{Related Work}
\label{sec:related}

\subsection{Communication Lower Bounds}

The pioneering work of Hong and Kung \cite{HK81} introduced a framework for communication analysis in the sequential model. 
Using the red-blue pebble game, Hong and Kung derived lower bounds on the number of words that must be communicated when performing a class of algorithms including conventional matrix multiplication.
Irony \emph{et al.} \cite{ITT04} extended Hong and Kung's results for matrix multiplication to the parallel case using a segmentation argument that we will follow.  
Ballard \emph{et al.} \cite{BCDH+14} extended communication lower bounds from matrix multiplication algorithms to algorithms for any linear algebra computations that can be written as three nested loop (3NL) computations.  
Smith and van de Geijn \cite{SvdG17} tightened the constants in the lower bounds given by Irony \emph{et al.} and Ballard \emph{et al.} by changing the operations to scalar fused multiply-adds, optimizing the segment length, and exploiting a bound on the sum (rather than the max) of the data accessed from each array.  
Additionally, memory-independent bounds were given by Ballard \emph{et al.} \cite{BDHLS12-SS} to determine the ranges where perfect strong scaling can be achieved. 
Demmel \emph{et al.} \cite{DE+13} considered how memory-independent bounds must change to remain tight for rectangular matrix multiplication with one, two, or three large dimensions.
Finally, Christ \emph{et al.} \cite{CDKSY13} extended the generality of 3NL computations to prove lower bounds for more arbitrary loop nests: their results apply to our definition of MTTKRP.

\subsection{Algorithms for MTTKRP}

The most straightforward sequential algorithm for MTTKRP, when the tensor is dense, involves permuting the tensor to achieve a column- or row-major matricization, forming the Khatri-Rao product explicitly, and then multiplying these two matrices \cite{BK07}.
Note that this approach violates the assumption in \cref{def:MTTKRP} that the \mults\ are performed atomically.
An alternative approach avoids the explicit permutation of the tensor and performs the MTTKRP in two steps, the first involving a matrix-matrix multiplication and the second involving a sequence of matrix-vector multiplications \cite{PTC13}.
This approach also violates the atomicity assumption.
The two-step approach is particularly advantageous when the MTTKRP is to be performed in each mode, like in the CP-ALS or other gradient-based algorithms, as intermediate quantities can be re-used across modes.

In the case of distributed-memory parallel algorithms for MTTKRP, there have been many efforts to improve performance for sparse tensors \cite{BK07,CV14,KU15,SK16} in the context of the CP-ALS algorithm.
In particular, Smith and Karypis \cite{SK16} describe a ``medium-grained'' parallelization scheme that is designed for sparse tensors but can be applied to dense tensors.
Indeed, Liavas \emph{et al.} \cite{LK+17} apply the preceding approach to dense 3-way tensors in computing CP decompositions with non-negativity constraints.
Aggour and Yenner \cite{AY16} also parallel MTTKRP for dense tensors, using a scheme that parallelizes over only the largest dimension of a 3-way tensor.

\section{Lower Bounds}
\label{sec:LBs}

\subsection{Preliminary Lemmas}
\label{sec:lb-prelim}

In this section we state four lemmas that will be useful in our main results.
\Cref{lem:HBL} is an inequality that generalizes the Loomis-Whitney inequality \cite{LW49}, which has been used in proving communication lower bounds for matrix computations \cite{ITT04,BCDH+14}.
\Cref{lem:linProg} provides the solution to a particular linear program that appears in our lower bound proofs.
\Cref{lem:LMprod,lem:LMsum} give solutions to nonlinear optimization problems that appear in later proofs.

The following result concerning H\"older-Brascamp-Lieb-type multilinear inequalities appears in a more general form in \cite[Proposition~7.1]{BCCT10}; a simpler proof for our special case is given in \cite[Theorem~6.6]{CDKSY13}.
\begin{lemma}
\label{lem:HBL}
Consider any positive integers $d$ and $m$ and any $m$ projections $\phi_j \from \integers^d \to \integers^{d_j}$ ($d_j \le d$), each of which extracts $d_j$ coordinates $S_j \subseteq [d]$ and forgets the $d-d_j$ others.
Define 
\[ \PP = \big\{ \V{s} \in [0,1]^m : \M{\Delta} \cdot \V{s} \ge \V{1} \big\}\text,\]
where the $d\times m$ matrix $\M{\Delta}$ has entries
\[ \M{\Delta}_{i,j} = \begin{cases} 1 & i \in S_j \\ 0 & i \not\in S_j \end{cases} \text. \]
If $\V{s} \in \PP$, then for all $E \subseteq \integers^d$, 
\[ |E| \le \prod_{j \in [m]} |\phi_j(E)|^{s_j} \text.\] 
\end{lemma}

\begin{lemma}\label{lem:linProg}
The solution of the linear program
\begin{equation}
\label{eq:LP}
\min \V{1}^T\V{s} \quad \text{subject to} \quad \M{\Delta} \cdot \V{s} \ge \V{1} \text{ and } \V{s} \geq 0,
\end{equation}
where
\begin{equation*}
\M{\Delta} = \begin{pmatrix} \M{I}_{N\times N} & \V{1}_{N\times 1} \\ \V{1}_{1\times N} & 0 \end{pmatrix}\text,
\end{equation*}
is $\V{s}^* = (1/N,\ldots,1/N,1{-}1/N)^T$ with $\V{1}^T\V{s}^*=2{-}1/N$.
\end{lemma}
\begin{IEEEproof}
The dual linear program is 
\begin{equation*}
\max \V{1}^T\V{t} \quad \text{subject to} \quad \M{\Delta}^T \cdot \V{t} \leq \V{1} \text{ and } \V{t} \geq 0.
\end{equation*}
Note that $\V{t}^*=\V{s}^*$ is feasible, and $\V{1}^T\V{t}^* = \V{1}^T\V{s}^*$, so $\V{s}^*$ is a solution of the primal by linear duality.
\end{IEEEproof}

\begin{lemma}
\label{lem:LMprod} 
Given $\V{s}> \V{0}$, the optimization problem
\begin{equation*}
\max_{\V{x}\geq\V{0}} \prod_{i\in[m]} x_i^{s_i}  \quad \text{subject to} \quad \sum_{i\in[m]} x_i \leq c
\end{equation*}
yields the maximum value
\begin{equation*}
\label{eq:maxProd}
c^{\sum_i s_i}\prod_{j\in[m]} \left(\frac{s_j}{\sum_i s_i}\right)^{s_j}.
\end{equation*}
\end{lemma}
\begin{IEEEproof}
Without loss of generality, we may tighten our condition on the sum to be equality.  
If $\sum_i x_i < c$, we can increase one of the $x_i$ which would increase the product because all $x_i \geq 0$. 
Therefore the maximum product is achieved with equality in the constraint on the sum.  

We use Lagrange multipliers to find the maximum in terms of the exponents given by $\V{s}$.  
Our Lagrangian is
\begin{equation*}
\mathcal{L}(x_1,\ldots,x_m,\lambda) = x_1^{s_1}\cdots x_m^{s_m}-\lambda(x_1+\cdots+x_m-c),
\end{equation*}
which has partial derivatives
\begin{align*}
\frac{\partial\mathcal{L}}{\partial x_j} &= s_jx_j^{s_j-1}\prod_{i\neq j}x_i^{s_i}-\lambda, \\
\frac{\partial\mathcal{L}}{\partial\lambda} &= c-\sum_j x_j.
\end{align*}
Setting $\frac{\partial\mathcal{L}}{\partial x_j}=0$ for each $j$, we have for all $j\neq i$,
\begin{equation*}
s_jx_j^{s_j-1}\prod_{k\neq j}x_k^{s_k} = s_ix_i^{s_i-1}\prod_{k\neq i}x_k^{s_k}, 
\end{equation*}
or \( x_j = \frac{s_j}{s_i}x_i\).
Setting $\frac{\partial\mathcal{L}}{\partial \lambda}=0$, we have
\begin{equation*}
c = \sum_{i=1}^m x_i = \sum_{i\in[m]} \frac{s_i}{s_j}x_j, 
\end{equation*}
or \( x_j = \frac{c s_j}{\sum_i s_i} \) for each $j$, which implies
\begin{equation*}
\max\prod_{j\in[m]} x_j^{s_j} = \prod_{j\in[m]}\left(\frac{c s_j}{\sum_i s_i}\right)^{s_j} = c^{\sum_i s_i}\prod_{j\in[m]}\left(\frac{s_j}{\sum_i s_i}\right)^{s_j}. \\
\end{equation*}
\end{IEEEproof}

\begin{lemma}
\label{lem:LMsum} 
For any $\V{s}\geq\V{0}$, the optimization problem
\begin{equation*}
\min_{\V{x}\geq0} \sum_{i\in[m]} x_i \quad \text{subject to} \quad \prod_{i\in[m]} x_i^{s_i} \geq c
\end{equation*}
yields the minimum value
\begin{equation*}
\left(\frac{c}{\prod_i s_i^{s_i}}\right)^{1/\sum_i s_i}\sum_{i\in[m]} s_i.
\end{equation*}
\end{lemma}
\begin{IEEEproof}
As in the proof of \cref{lem:LMprod}, we note that our constraint can be changed to an equality constraint.  
If $\prod_{i\in[m]} x_i>c$, then we may decrease at least one $x_i$ and still have the constraint hold.  
This would decrease the sum, therefore the minimum sum must occur when there is equality in the constraint.  

As before, we use Lagrange multipliers to solve the optimization problem.  
Our Lagrangian is
\begin{equation*}
\mathcal{L}(x_1,\dots,x_m,\lambda) = \sum_{i\in[m]} x_i - \lambda\left(\prod_{i\in[m]} x_i^{s_i} - c\right)
\end{equation*}
and has partial derivatives
\begin{align*}
\frac{\partial\mathcal{L}}{\partial x_i} &= 1 - \lambda s_i x_i^{s_i-1}\prod_{j\neq i} x_j^{s_j}, \\
\frac{\partial\mathcal{L}}{\partial \lambda} &= c - \prod_{i\in[m]} x_i^{s_i}.
\end{align*}
Setting the partial derivatives with respect to $x_i$ to zero we can again derive that for all pairs $j\neq i$, $x_j = \frac{s_j}{s_i}x_i$.  
Additionally when $\frac{\partial\mathcal{L}}{\partial\lambda} = 0$,
\begin{equation*}
c=\prod_{i\in[m]} x_i^{s_i} = \left(\frac{x_j}{s_j}\right)^{\sum_i s_i}\prod_{i\in[m]} s_i^{s_i}, 
\end{equation*}
or $x_j = s_j \left(\frac{c}{\prod_i s_i^{s_i}}\right)^{1/\sum_i s_i}$ for each $j$.
So
\begin{equation*}
\min_{\V{x}\ge 0} \sum_{j\in[m]} x_j = \left(\frac{c}{\prod_i s_i^{s_i}}\right)^{1/\sum_i s_i}\sum_{i\in[m]} s_i.
\end{equation*}
\end{IEEEproof}

\subsection{Memory-Dependent Lower Bounds}
\label{sec:lb-mem-dep}

We first prove \cref{thm:seq_LB}, a lower bound for the sequential model that depends on the fast memory size $M$.
The proof uses the structure of previous matrix computation lower bound proofs \cite{ITT04,BCDH+14}.
However, to address MTTKRP, it uses a H\"{o}lder-Brascamp-Lieb-type inquality (\cref{lem:HBL}) as has been done for more general computations \cite{CDKSY13}.
It also borrows another technique involving \cref{lem:LMprod} that has been used to tighten the constant of the matrix multiplication bound \cite{SvdG17}, though the technique improves our bound by more than a constant. 
\Cref{thm:seq_LB} implies \cref{cor:par_memind_LB}, a similar memory-independent bound for the parallel model, where $M$ corresponds to the size of the local memory.
We also state an immediate lower bound result for the sequential case (\cref{fct:trivial-lb}) based on the size of the input and output data.

\begin{theorem}
\label{thm:seq_LB}
Any sequential MTTKRP algorithm involves at least
\begin{equation}
\label{eq:seq_LB}
\frac{1}{3^{2-1/N}}\frac{NIR}{M^{1-1/N}} - M
\end{equation}
loads and stores.
\end{theorem}

\begin{IEEEproof}
We break the stream of instructions that implement a MTTKRP algorithm into \emph{complete segments} each of which contains exactly $M$ loads and stores, except the last segment which may contain less than $M$ loads and stores (\emph{incomplete}).
We will determine an upper bound on the number of elements of all arrays $\X$, $\Mn{B}{n}$, or $\Mn{A}{k}$ that can be accessed during a segment, then use \cref{lem:HBL} to bound the number of loop iterations that can be evaluated during a segment.  
We use this upper bound to generate a lower bound for the number of complete segments, from which we generate the lower bound on the communication for any MTTKRP algorithm.

We begin by considering elements of $\Mn{B}{n}$, the factor matrix that is being computed.  
We consider an element of $\Mn{B}{n}$ \emph{live} during the segment if it accumulates the result of one or more \mults\ during that segment.  
Any element of $\Mn{B}{n}$ that is live during the segment must either remain in fast memory at the end of the segment or have been stored into slow memory by the end of the segment.  
At the end of the segment there can be at most $M$ live elements of $\Mn{B}{n}$ that remain in fast memory.  
Let $S$ be the number of live elements of $\Mn{B}{n}$ that were stored during the segment.  
Now, consider input elements of $\X$ and $\Mn{A}{k}$ that are used as arguments for one or more \mults\ during the segment.  
These elements must have been in fast memory at the start of the segment or loaded into fast memory during the segment.  
The total number of input elements that are in fast memory at the start of segment is at most $M$, and the total number of input elements that can be loaded during the segment is $M-S$.  
Thus the total number elements from all arrays that an algorithm can access during the segment is at most $3M$.

If $F$ is the subset of the iteration space $\I = [I_1] \times \cdots \times [I_N] \times [R]$ evaluated during the segment, then $\phi_j (F)$ corresponds to the set of entries of the $j$-th array that are accessed during the segment. 
Thus, 
\begin{equation*}
\sum_{j\in[m]} |\phi_j(F)| \leq 3M.
\end{equation*}
See \Cref{fig:projections} for an example set $F$ and its projections.

\newcommand{\tdim}{16}
\newcommand{\rdim}{5}
\newcommand{\face}{\draw[black] (0,0) rectangle (\tdim,\tdim);}
\newcommand{\lbscale}{.0825}

\newcommand{\drawtensor}{
\begin{scope}[canvas is yz plane at x=0,rotate=-90]
	\face
\end{scope}
\begin{scope}[canvas is yx plane at z=0,yscale=-1,rotate=0]
	\face
\end{scope}
\begin{scope}[canvas is zx plane at y=(\tdim),rotate=180]
	\face
\end{scope}
}

\newcommand{\drawfacmat}{
\draw[black] (0,0) rectangle (\rdim,\tdim);
\draw[dotted] (\rdim/4,0) -- (\rdim/4,\tdim);
\draw[dotted] (2*\rdim/4,0) -- (2*\rdim/4,\tdim);
\draw[dotted] (3*\rdim/4,0) -- (3*\rdim/4,\tdim);
}

\newcommand{\ia}{5}
\newcommand{\ja}{1}
\newcommand{\ka}{1}
\newcommand{\ra}{1}
\newcommand{\ib}{3}
\newcommand{\jb}{3}
\newcommand{\kb}{15}
\newcommand{\rb}{1}
\newcommand{\ic}{7}
\newcommand{\jc}{10}
\newcommand{\kc}{2}
\newcommand{\rc}{2}
\newcommand{\id}{4}
\newcommand{\jd}{14}
\newcommand{\kd}{11}
\newcommand{\rd}{3}
\newcommand{\ie}{11}
\newcommand{\je}{2}
\newcommand{\ke}{2}
\newcommand{\re}{4}
\newcommand{\iif}{14}
\newcommand{\jf}{14}
\newcommand{\kf}{14}
\newcommand{\rf}{4}

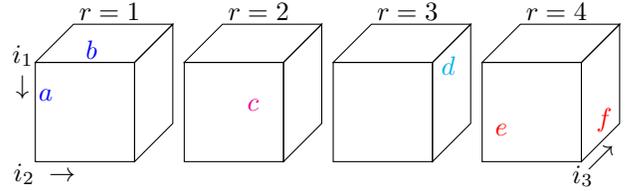
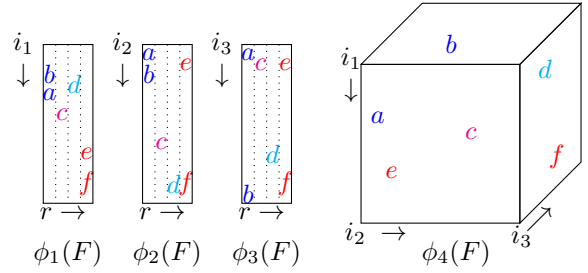
\begin{figure}
\centering
\subfloat[\label{fig:projections:it} Example subset $F$ of 4-way iteration space.  
The subset $F$ consists of the six coordinates $a$ (5,1,1,1), $b$ (3,3,15,1), $c$ (7,10,2,2), $d$ (4,14,11,3), $e$ (11,2,2,4), and $f$ (14,14,14,4), which are color coded by their last index.]
{\begin{tikzpicture}[scale=\lbscale]

\begin{scope}[color=blue]
	\begin{scope}[canvas is yx plane at z=-\ka+.5,yscale=-1,rotate=180,shift={(-\tdim+.5,-\tdim+.5)}]
		\node (a) at (\ia,\ja) {$a$};
	\end{scope}
	\begin{scope}[canvas is yx plane at z=-\kb+.5,yscale=-1,rotate=180,shift={(-\tdim+.5,-\tdim+.5)}]
		\node (b) at (\ib,\jb) {$b$};
	\end{scope}
	\drawtensor
	\node[color=black] (i1) at (-\tdim-2,\tdim-1.5) {$\begin{array}{c} i_1 \\ \downarrow \end{array}$};
	\node[color=black] (i2) at (-\tdim+1.5,-2) {$i_2 \; \rightarrow$};
	\node[color=black] (r1) at (-\tdim/4,1.5*\tdim+.25) {$r=1$};
\end{scope}

\begin{scope}[color=magenta,shift={(1.5*\tdim,0)}]
	\begin{scope}[canvas is yx plane at z=-\kc+.5,yscale=-1,rotate=180,shift={(-\tdim+.5,-\tdim+.5)}]
		\node (c) at (\ic,\jc) {$c$};
	\end{scope}
	\drawtensor
	\node[color=black] (r2) at (-\tdim/4,1.5*\tdim+.25) {$r=2$};
\end{scope}

\begin{scope}[color=cyan,shift={(3*\tdim,0)}]
	\begin{scope}[canvas is yx plane at z=-\kd+.5,yscale=-1,rotate=180,shift={(-\tdim+.5,-\tdim+.5)}]
		\node (d) at (\id,\jd) {$d$};
	\end{scope}
	\drawtensor
	\node[color=black] (r3) at (-\tdim/4,1.5*\tdim+.25) {$r=3$};
\end{scope}

\begin{scope}[color=red,shift={(4.5*\tdim,0)}]
	\begin{scope}[canvas is yx plane at z=-\ke+.5,yscale=-1,rotate=180,shift={(-\tdim+.5,-\tdim+.5)}]
		\node (e) at (\ie,\je) {$e$};
	\end{scope}
	\begin{scope}[canvas is yx plane at z=-\kf+.5,yscale=-1,rotate=180,shift={(-\tdim+.5,-\tdim+.5)}]
		\node (f) at (\iif,\jf) {$f$};
	\end{scope}
	\drawtensor
	\begin{scope}[canvas is yz plane at x=0,rotate=-90]
		\node[color=black] (i3) at (.5,-2.5) {$i_3$};
		\node[color=black] (arrow3) at (8,-2.5) {$\nearrow$};
	\end{scope}
	\node[color=black] (r4) at (-\tdim/4,1.5*\tdim+.25) {$r=4$};
\end{scope}

\end{tikzpicture}} \\
\subfloat[\label{fig:projections:arr} Projections of $F$ onto data arrays (2-way factor matrices and 3-way tensor).  
For example, the set $\phi_2(F)$ consists of the six coordinates $a$ (1,1), $b$ (3,1), $c$ (10,2), $d$ (14,3), $e$ (2,4), and $f$ (14,4).]
{\begin{tikzpicture}[scale=\lbscale*1.6]

\begin{scope}
	\begin{scope}[yscale=-1]
		\node[color=blue] (a1) at (\ra*\rdim/4-\rdim/8,\ia) {$a$};
		\node[color=blue] (b1) at (\rb*\rdim/4-\rdim/8,\ib) {$b$};
		\node[color=magenta] (c1) at (\rc*\rdim/4-\rdim/8,\ic) {$c$};
		\node[color=cyan] (d1) at (\rd*\rdim/4-\rdim/8,\id) {$d$};
		\node[color=red] (e1) at (\re*\rdim/4-\rdim/8,\ie) {$e$};
		\node[color=red] (f1) at (\rf*\rdim/4-\rdim/8,\iif) {$f$};
		\drawfacmat
		\node[color=black] (i1m) at (-2,1.5) {$\begin{array}{c} i_1 \\ \downarrow \end{array}$};
		\node[color=black] (r1) at (2,\tdim+1) {$r \rightarrow$};
		\node (phi4) at (\rdim/2,1.125*\tdim+3) {$\phi_1(F)$};
	\end{scope}
	\begin{scope}[yscale=-1,shift={(2*\rdim,0)}]
		\node[color=blue] (a1) at (\ra*\rdim/4-\rdim/8,\ja) {$a$};
		\node[color=blue] (b1) at (\rb*\rdim/4-\rdim/8,\jb) {$b$};
		\node[color=magenta] (c1) at (\rc*\rdim/4-\rdim/8,\jc) {$c$};
		\node[color=cyan] (d1) at (\rd*\rdim/4-\rdim/8,\jd) {$d$};
		\node[color=red] (e1) at (\re*\rdim/4-\rdim/8,\je) {$e$};
		\node[color=red] (f1) at (\rf*\rdim/4-\rdim/8,\jf) {$f$};
		\drawfacmat
		\node[color=black] (i2m) at (-2,1.5) {$\begin{array}{c} i_2 \\ \downarrow \end{array}$};
		\node[color=black] (r2) at (2,\tdim+1) {$r \rightarrow$};
		\node (phi4) at (\rdim/2,1.125*\tdim+3) {$\phi_2(F)$};
	\end{scope}
	\begin{scope}[yscale=-1,shift={(4*\rdim,0)}]
		\node[color=blue] (a1) at (\ra*\rdim/4-\rdim/8,\ka) {$a$};
		\node[color=blue] (b1) at (\rb*\rdim/4-\rdim/8,\kb) {$b$};
		\node[color=magenta] (c1) at (\rc*\rdim/4-\rdim/8,\kc) {$c$};
		\node[color=cyan] (d1) at (\rd*\rdim/4-\rdim/8,\kd) {$d$};
		\node[color=red] (e1) at (\re*\rdim/4-\rdim/8,\ke) {$e$};
		\node[color=red] (f1) at (\rf*\rdim/4-\rdim/8,\kf) {$f$};
		\drawfacmat
		\node[color=black] (i3m) at (-2,1.5) {$\begin{array}{c} i_3 \\ \downarrow \end{array}$};
		\node[color=black] (r3) at (2,\tdim+1) {$r \rightarrow$};
		\node (phi4) at (\rdim/2,1.125*\tdim+3) {$\phi_3(F)$};
	\end{scope}
\end{scope}

\begin{scope}[shift={(3*\tdim,-1.125*\tdim)}]
	\begin{scope}[canvas is yx plane at z=-\ka+.5,yscale=-1,rotate=180,shift={(-\tdim+.5,-\tdim+.5)}]
		\node[color=blue] (at) at (\ia,\ja) {$a$};
	\end{scope}
	\begin{scope}[canvas is yx plane at z=-\kb+.5,yscale=-1,rotate=180,shift={(-\tdim+.5,-\tdim+.5)}]
		\node[color=blue] (bt) at (\ib,\jb) {$b$};
	\end{scope}
	\begin{scope}[canvas is yx plane at z=-\kc+.5,yscale=-1,rotate=180,shift={(-\tdim+.5,-\tdim+.5)}]
		\node[color=magenta] (ct) at (\ic,\jc) {$c$};
	\end{scope}
	\begin{scope}[canvas is yx plane at z=-\kd+.5,yscale=-1,rotate=180,shift={(-\tdim+.5,-\tdim+.5)}]
		\node[color=cyan] (dt) at (\id,\jd) {$d$};
	\end{scope}
	\begin{scope}[canvas is yx plane at z=-\ke+.5,yscale=-1,rotate=180,shift={(-\tdim+.5,-\tdim+.5)}]
		\node[color=red] (et) at (\ie,\je) {$e$};
	\end{scope}
	\begin{scope}[canvas is yx plane at z=-\kf+.5,yscale=-1,rotate=180,shift={(-\tdim+.5,-\tdim+.5)}]
		\node[color=red] (ft) at (\iif,\jf) {$f$};
	\end{scope}
	\drawtensor
	\node[color=black] (i1) at (-\tdim-1,\tdim-1) {$\begin{array}{c} i_1 \\ \downarrow \end{array}$};
	\node[color=black] (i2) at (-\tdim+1.5,-1) {$i_2 \; \rightarrow$};
	\begin{scope}[canvas is yz plane at x=0,rotate=-90]
		\node[color=black] (i3) at (.5,-1.5) {$i_3$};
		\node[color=black] (arrow3) at (5,-1.5) {$\nearrow$};
	\end{scope}
	\node (phi4) at (-.4*\tdim,-3) {$\phi_4(F)$};
\end{scope}

\end{tikzpicture}}
\caption[fig:arrayaccess]{Example subset of computation and the data required to perform it, for $N=3$, $I_1=I_2=I_3=15$, and $R=4$.
\Cref{fig:projections:it} shows the iteration space and specifies six coordinates in the subset, where the coordinates correspond to \mults.
\Cref{fig:projections:arr} show the elements of the arrays that are involved in the computation, which are determined by projections of the coordinates.}
\label{fig:projections}
\end{figure}

To use \cref{lem:HBL} we first define the linear constraint matrix $\Delta$.  
For MTTKRP algorithms, the number of projections/arrays is $m = N{+}1$, corresponding to $N{-}1$ input factor matrices, one output factor matrix, and the input tensor.  
The depth of the nested loops is $d=N{+}1$, corresponding to one loop for each mode of the tensor and one loop over the rank of the factor matrices.  
The first $N$ projections (rows) correspond to the input and output factor matrices, and the last projection corresponds to the input tensor.
The first $N$ indices (columns) are $i_1,\ldots,i_N$, and the last index is $r$.
So we have
\begin{equation*}
\M{\Delta} = \begin{pmatrix} \M{I}_{N\times N} & \V{1}_{N\times 1} \\ \V{1}_{1\times N} & 0 \end{pmatrix}\text.
\end{equation*}
%
%
By \cref{lem:HBL}, for any $\V{s} \in \PP$,
\begin{equation*}
|F| \leq \prod_{j\in[m]} |\phi_j(F)|^{s_j}
\end{equation*}

Substituting $|\phi_j(F)|$ for $x_j$ and $3M$ as the constant $c$ in the constraint of \cref{lem:LMprod}, we see that for any $\V{s}\in\PP$,
\begin{align*}
\prod_{j\in[m]}|\phi_j(F)|^{s_j} \leq (3M)^{\sum_j s_j}\prod_{j\in[m]}\left(\frac{s_j}{\sum_i s_i}\right)^{s_j}\text{.}
\end{align*}
In order to obtain the tightest lower bound possible, we wish to choose the $\V{s}\in\PP$ that minimizes the left hand side of the preceding inequality.
Short of that, we can choose to minimize only the first factor $(3M)^{\sum_j s_j}$, which corresponds to solving the linear program \cref{eq:LP}.
By \cref{lem:linProg}, the exponent is minimized by $2{-}1/N$ with $\V{s}^* = (1/N,\ldots,1/N,1{-}1/N)^T$.  
Note that 
\begin{align*}
&\prod_{j\in[m]}\left(\frac{s_j^*}{\sum_i s_i^*}\right)^{s_j^*} = \left(\frac{1-1/N}{2-1/N}\right)^{1-1/N}\prod_{j\in[N]}\left(\frac{1/N}{2-1/N}\right)^{1/N} \\
&\qquad= \left(\frac{1}{2-1/N}\right)^{2-1/N}(1-1/N)^{1-1/N}\prod_{j\in[N]} (1/N)^{1/N} \\
&\qquad\leq 1/N.
\end{align*}
Thus $|F|\le (3M)^{2-1/N}/N$ gives an upper bound on the number of \mults\ that can be performed in a segment with exactly $M$ loads and stores.

Because $|\I| = IR$ there are at least
\begin{equation*}
\left\lfloor\frac{IR}{(3M)^{2-1/N}/N}\right\rfloor
\end{equation*}
complete segments.
Each segment loads and stores $M$ words, thus there are at least
\begin{equation*}
M\cdot\left\lfloor\frac{NIR}{(3M)^{2-1/N}}\right\rfloor
\end{equation*}
loads and stores.
\end{IEEEproof}

\begin{corollary}
\label{cor:par_memind_LB}
Any parallel MTTKRP algorithm involves at least
\begin{equation*}
\frac{1}{3^{2-1/N}}\frac{NIR}{PM^{1-1/N}} - M
\end{equation*}
sends and receives.
\end{corollary}
\begin{IEEEproof}
Since some processor must be associated with at least $|\mathcal{I}|/P = IR/P$ loop iterations, we can apply \cref{thm:seq_LB} to the computation performed by that processor.
\end{IEEEproof}

The following additional lower bound for the sequential case is based on the observation that to perform the MTTKRP, the algorithm must access all of the input and output data.
Note that the fast memory could be full of useful data at the beginning and end of the computation.

\begin{fact}
\label{fct:trivial-lb}
Any sequential MTTKRP algorithm must perform at least
\begin{equation}
\label{eq:trivial-lb}
I + \sum_{k\in [N]} I_kR - 2M
\end{equation}
loads and stores.
\end{fact}

\subsection{Memory-Independent Lower Bounds}
\label{sec:lb-mem-indep}

In this section, we prove bounds that do not depend on the fast or local memory size $M$.
These bounds focus on the parallel case.
The structures of the proofs follow previous work \cite{BDHLS12-SS,DE+13}, but again we combine a technique used in the context of matrix multiplication \cite{SvdG17} (involving \cref{lem:LMprod}) to tighten the bounds.
\Cref{thm:MemIndepLB,thm:rectMemIndepLB} establish separate lower bounds under the same assumptions on the parallelization and data distribution.
We prove both because either can be the tightest lower bound, depending on relative sizes of the parameters.
To show how the bounds simplify and compare for a particular case, we consider tensors with all dimensions the same ($I_k=I^{1/N}$ for all $k$) and state \cref{cor:par_LB}.

%
\begin{theorem}
\label{thm:MemIndepLB}
In any parallel MTTKRP algorithm where each processor initially and finally owns at most $\delta \sum_k I_kR/P$ factor matrix entries and at most $\gamma I/P$ tensor entries, $\gamma,\delta\geq 1$, some processor performs at least
\begin{equation}
\label{eq:MemIndepLB}
2\left(\frac{NIR}{P}\right)^{\frac{N}{2N-1}}-\gamma\frac{I}{P}-\delta\sum_{k\in[N]}\frac{I_kR}{P}
\end{equation}
sends and receives.
\end{theorem}
\begin{IEEEproof}
We follow the argument given by Ballard \emph{et al.} \cite[Lemma 2.3]{BDHLS12-SS}.
Some processor $p$ must evaluate at least $|\I|/P=IR/P$ loop iterations.
Let $F$ be the set of loop iterations associated with the \mults\ performed by that processor.  
Then using $|\phi_j(F)|$ as before we have that the number of sends and receives performed by that processor must be at least
\begin{equation*} 
 \sum_{j\in[N+1]}|\phi_j(F)|-\gamma I/P-\delta \sum_{k\in[N]} I_kR/P \text,
\end{equation*}
where the first sum is the size of the data the processor must access to evaluate its loop iterations and the negative terms correspond to the useful data that may be in its local memory at the start and end of the computation.
From \cref{lem:HBL}, we can bound the size of $F$ in terms of the sizes of the projections:
\begin{equation*}
|F| \le \prod_{j\in[N+1]} |\phi_j(F)|^{s_j}
\end{equation*}
for any $\V s$ in $\PP$. 
Using $\V{s}^*=(1/N,\ldots,1/N,1{-}1/N)$ as before, and substituting $|\phi_j(F)|$ for $x_j$ and $IR/P$ as the constant $c$, \cref{lem:LMsum} gives
\begin{align*}
\sum_{j\in[N+1]} |\phi_j(F)| &\ge \left(\frac{IR/P}{\prod_{j\in[N+1]} s_j^{s_j}}\right)^{\frac{N}{2N-1}}(2-1/N) \\
&\ge 2\left(\frac{NIR}{P}\right)^{\frac{N}{2N-1}}\text.
\end{align*}
\end{IEEEproof}

%
%
\begin{theorem}
\label{thm:rectMemIndepLB}
In any parallel MTTKRP algorithm where each processor initially and finally owns at most $\delta \sum_k I_kR/P$ factor matrix entries and at most $\gamma I/P$ tensor entries, $\gamma,\delta\geq 1$, some processor performs at least
\begin{equation}
\label{eq:rectMemIndepLB}
\min\left(\sqrt{\frac{2}{3\gamma}}NR\left(\frac{I}{P}\right)^{1/N}-\delta\sum_{j\in[N]} \frac{I_jR}{P},\ \frac{\gamma I}{2 P}\right)
\end{equation}
sends and receives.
\end{theorem}
\begin{IEEEproof}
We follow the argument given by Demmel \emph{et al.}~\cite[Section II.B.2]{DE+13}.  
As before, $F$ is the set of loop iterations evaluated by a processor that computes at least $IR/P$ \mults.  
By \cref{lem:HBL} with $\V{s}^*=(1/N,\dots,1/N,1{-}1/N)^T$, we have
\begin{equation}
\label{eq:carma}
\frac{IR}{P} \leq |\phi_{N+1}(F)|^{\frac{N-1}{N}}\prod_{j\in[N]} |\phi_j(F)|^{1/N}.
\end{equation}

We consider two cases based on $|\phi_{N+1}(F)|$, the number of tensor entries accessed by the processor.  
Suppose that $|\phi_{N+1}(F)| \geq \frac{3\gamma I}{2P}$.  
By our assumption of load balanced data distribution, the processor must read at least $\frac{\gamma I}{2P}$ elements of $\X$ to perform its computations.
Now consider the case when $|\phi_{N+1}(F)| < \frac{3\gamma I}{2P}$.  
Replacing $|\phi_{N+1}(F)|$ with $\frac{3\gamma I}{2P}$ in the right hand side of \cref{eq:carma} and rearranging, we have
\begin{equation*}
\prod_{j\in[N]} |\phi_j(F)| \geq \left(\frac{2}{3\gamma}\right)^{N-1}\frac{IR^N}{P}.
\end{equation*}
By \cref{lem:LMsum}, we know that $\sum_{j\in[N]} |\phi_j(F)|$ is minimized subject to this constraint on the product when 
\begin{equation*}
|\phi_j(F)| = \left(\frac{2}{3\gamma}\right)^{\frac{N-1}{N}}\left(\frac{I}{P}\right)^{1/N}R 
\end{equation*}
Given that the factor matrices are load balanced up to a factor of $\delta$, we see that some processor performs at least
\begin{multline*}
\sum_{j\in[N]} |\phi_j(F)|-\delta \sum_{j\in[N]} \frac{I_jR}{P} 
\\
\geq N\left(\frac{2}{3\gamma}\right)^{\frac{N-1}{N}}\left(\frac{I}{P}\right)^{1/N}R - \delta\sum_{j\in[N]} \frac{I_j R}{P}
\end{multline*}
sends and receives.

Because the number of tensor entries the processor must access may be bigger or smaller than $\frac{3\gamma I}{2P}$, the lower bound is the minimum of the two cases.
\end{IEEEproof}

\begin{corollary}
\label{cor:par_LB}
Any parallel MTTKRP algorithm involving a tensor with $I_k=I^{1/N}$ for all $k$ and that starts with one copy of the inputs evenly distributed across processors and ends with one copy of the output evenly distributed across processors involves at least\begin{equation*}
\Omega\left( \left( \frac{NIR}{P} \right)^{\frac{N}{2N-1}} + NR\left(\frac{I}{P}\right)^{1/N} \right)
\end{equation*}
sends and receives.
\end{corollary}

\begin{IEEEproof}
Under these assumptions, both \cref{thm:MemIndepLB,thm:rectMemIndepLB} apply.
Given that $I_k = I^{1/N}$, we can simplify the bound from  \cref{thm:MemIndepLB} to
\begin{equation}
\label{eq:parLB1}
\Omega\left(\left(\frac{NIR}{P}\right)^\frac{N}{2N-1}-\frac IP - \frac{NI^{1/N}R}{P} \right),
\end{equation}
and we can simplify the bound from \cref{thm:rectMemIndepLB} to
\begin{equation}
\label{eq:parLB2}
\Omega\left(\min\left(NR\left(\frac{I}{P}\right)^{1/N},\ \frac IP \right)\right),
\end{equation}
assuming $P>1$.

We now consider two cases.
Suppose $NR\geq (I/P)^{1-1/N}$.
This implies that $(NIR/P)^{\frac{N}{2N-1}}$ dominates $I/P$, which implies that \cref{eq:parLB1} dominates \cref{eq:parLB2} and simplifies to 
\[\Omega\left(\left(\frac{NIR}{P}\right)^\frac{N}{2N-1} - \frac{NI^{1/N}R}{P} \right).\]
Again, $NR\geq (I/P)^{1-1/N}$ implies that
\[\left(\frac{NIR}{P}\right)^{\frac{N}{2N-1}} \geq NR\left(\frac IP\right)^{1/N} \geq \frac{NI^{1/N}R}{P},\]
and the bound
\[\Omega\left(\left(\frac{NIR}{P}\right)^{\frac{N}{2N-1}} \right)\]
applies.

Suppose $NR\leq (I/P)^{1-1/N}$.
This implies that \cref{eq:parLB1} degenerates to a negative bound and \cref{eq:parLB2} simplifies to 
\[ \Omega\left(NR\left(\frac IP\right)^{1/N} \right). \]

The two bounds apply in separate cases, but because the first bound dominates when $NR$ is larger than the threshold $(I/P)^{1-1/N}$ and the second bound dominates when $NR$ is smaller than the threshold, we can write the overall bound as a sum of the two bounds, as stated.
\end{IEEEproof}

\section{Algorithms}
\label{sec:algs}

\begin{algorithm}
\small
\caption{Sequential Unblocked Algorithm}
\label{alg:seq-unblocked}
\begin{algorithmic}[1]
\Function{$\Mn{B}{n}$ $=$ Seq-MTTKRP}{$\X$, $\{\Mn{A}{k}\}_{k \in [N] \setminus\{n\}}$, $n$}
\For{$i_1\gets1$ \textbf{to} $I_1$}
	\ForDots
		\For{$i_N\gets1$ \textbf{to} $I_N$}
			\State load $\X(i_1,\,\ldots,\,i_N)$
			\For{$r\gets1$ \textbf{to} $R$}
				\State load $\Mn{A}{k}(i_k,\,r)$ ($k \in [N] \setminus\{n\}$)
				\State load $\Mn{B}{n}(i_n,\,r)$ 
				\State 
					$\Mn{B}{n}(i_n,\,r) \gets 
					\Mn{B}{n}(i_n,\,r)+$
				\Statex \hfill $\X(i_1,\,\ldots,\,i_N) \cdot 
					\displaystyle\prod_{\mathclap{k \in [N] \setminus\{n\}}} \Mn{A}{k}(i_k,\,r)$
				\State store $\Mn{B}{n}(i_n,\,r)$
			\EndFor
		\EndFor
	\EndForDots
\EndFor
\EndFunction

\end{algorithmic}
\end{algorithm}

\begin{algorithm}
\small
\caption{Sequential Blocked Algorithm}
\label{alg:seq-blocked}
\begin{algorithmic}[1]
\Function{$\Mn{B}{n}$ $=$ Seq-Blocked-MTTKRP}{$\X$, $\{\Mn{A}{k}\}$, $n$, $b$}
\For{$j_1\gets1$ \textbf{to} $I_1$ \textbf{step} $b$}
	\ForDots
		\For{$j_N\gets1$ \textbf{to} $I_N$ \textbf{step} $b$}
			\State $J_k \gets \min(I_k,\,j_k+b-1)$ ($k \in [N]$)
			\State load block $\X(j_1{:}J_1,\,\ldots,\,j_N{:}J_N)$
			\For{$r \gets 1$ \textbf{to} $R$}
				\State load vectors $\Mn{A}{k}(j_k{:}J_k,\,r)$ ($k \in [N] \setminus\{n\}$)
				\State load vector  $\Mn{B}{n}(j_n{:}J_n,\,r)$ 
				\For{$i_1\gets j_1$ \textbf{to} $J_1$}
					\ForDots
						\For{$i_N\gets j_N$ \textbf{to} $J_N$}
							\State 
								$\Mn{B}{n}(i_n,\,r) \gets 
								\Mn{B}{n}(i_n,\,r)+$
							\Statex \hfill $\X(i_1,\,\ldots,\,i_N) \cdot 
								\displaystyle\prod_{\mathclap{k \in [N] \setminus\{n\}}} \Mn{A}{k}(i_k,\,r)$
						\EndFor
					\EndForDots
				\EndFor
				\State store vector $\Mn{B}{n}(j_n{:}J_n,\,r)$
			\EndFor
		\EndFor
	\EndForDots
\EndFor
\EndFunction
\end{algorithmic}
\end{algorithm}

\subsection{Sequential Unblocked Algorithm}
\label{sec:impl-unblocked}

\Cref{alg:seq-unblocked} illustrates a sequential MTTKRP algorithm.
It makes no assumption on the fast memory size besides $M \ge N$ (necessary for \mults).
The communication cost of \cref{alg:seq-unblocked} is
\[ W \le I + IR(N+1)\text;\]
the two terms bound the numbers of tensor entry loads and factor-matrix entry loads/stores, respectively.

This counting neglects the possibility that inputs/outputs begin/end in fast memory.
For example, if $I+(I_1+\cdots+I_N)R\le M$, then $W=0$ is attained.

\subsection{Sequential Blocked Algorithm}
\label{sec:impl-seq-blocked}

\Cref{alg:seq-blocked} illustrates another sequential MTTKRP algorithm.
The iterations are performed in a different order, which potentially exposes more data reuse. 

\begin{figure}
\centering
\newcommand{\matdim}{4}
\newcommand{\mat}{\draw[black,shift={(-.5,-.5)}] (0,0) grid (\matdim,\matdim);}
\newcommand{\highlight}{gray!50}

\begin{tikzpicture}[x={(-0.5cm,-0.4cm)}, y={(1cm,0cm)}, z={(0cm,1cm)},every node/.append style={transform shape},scale=.75]

\begin{scope}[canvas is yz plane at x=.5,shift={(1.5,-1.5)}]
	\draw[dashed,fill=\highlight,shift={(0,0)}] (0,0) rectangle (-1,-1);
	\node[draw=none] at (-1.25,-.5) {\small $b$};
	\node[draw=none] at (-.5,-1.25) {\small $b$};
	\draw[fill=\highlight,shift={(-2.5,0)},xscale=.5] (0,0) rectangle (-1,-1);
	\draw[fill=\highlight,shift={(0,-2.5)},yscale=.5] (0,0) rectangle (-1,-1);
	\draw[densely dotted,thick,shift={(-2.8,0)}] (0,0) -- (0,-1);
	\draw[densely dotted,thick,shift={(0,-2.8)}] (0,0) -- (-1,0);
\end{scope}
\begin{scope}[canvas is zx plane at y=(\matdim-.5),rotate=-90,shift={(.5,-4)}]
	\draw[fill=\highlight,yscale=.5] (0,0) rectangle (-1,-1);
	\draw[densely dotted,thick,shift={(0,-.3)}] (0,0) -- (-1,0);
\end{scope}
\begin{scope}[canvas is zx plane at y=1.5,rotate=-90,shift={(-.5,-2.5)}]
	\draw[dashed,fill=\highlight] (0,0) rectangle (1,1);
	\node[draw=none] at (.75,-.2) {\small $b$};
\end{scope}
\begin{scope}[canvas is yx plane at z=-1.5,yscale=-1,rotate=0,shift={(.5,-.5)}]
	\draw (0,0) node {\LARGE $\cdot$};
	\draw[dashed,fill=\highlight] (0,0) rectangle (1,1);
\end{scope}

\begin{scope}[canvas is yz plane at x=.5,rotate=-90]
	\mat
\end{scope}
\begin{scope}[canvas is yx plane at z=.5,yscale=-1,rotate=0]
	\mat
\end{scope}
\begin{scope}[canvas is zx plane at y=(\matdim-.5),rotate=180]
	\mat
\end{scope}

\begin{scope}[canvas is yz plane at x=.5,shift={(1.5,-1.5)}]
	\draw[shift={(-2.5,0)},xscale=.5] (0,-2) grid (-1,2);
	\node[draw=none] at (-3.5,0) {$\Mn{A}{1}$};
	\draw[shift={(0,-2.5)},yscale=.5] (-2,0) grid (2,-1);
	\node[draw=none] at (0,-3.5) {$\Mn{B}{2}$};
\end{scope}
\begin{scope}[canvas is zx plane at y=(\matdim-.5),rotate=-90,shift={(1.5,-1.5)}]
	\draw[shift={(0,-2.5)},yscale=.5] (-2,0) grid (2,-1);
	\node[draw=none] at (0,-3.5) {$\Mn{A}{3}$};
\end{scope}
\end{tikzpicture}
\caption[fig:Seq.Block]{Sequential Blocked Algorithm for $N=3$ and $n=2$: subtensor $\X(j_1{:}J_1,j_2{:}J_2,j_3{:}J_3)$ is highlighted, and subcolumns $\Mn{A}{1}(j_1{:}J_1,\,r)$, $\Mn{B}{2}(j_2{:}J_2,\,r)$, $\Mn{A}{3}(j_3{:}J_3,\,r)$ are shown with dotted lines.}
\end{figure}
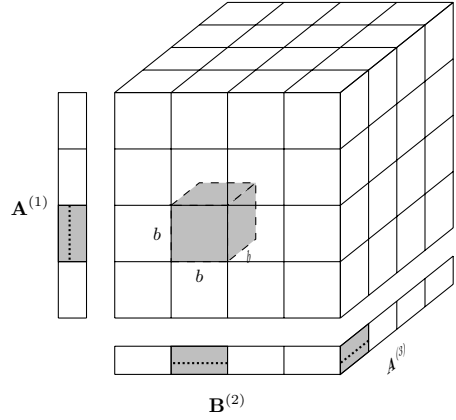

We control the blocking with the block size $b$. 
The code is correct for any positive integer $b$ satisfying
\begin{equation}
\label{eq:seq-b-constraint}
b^N + Nb \le M \text, 
\end{equation}
whence the communication cost is bounded above by
\begin{equation}
\label{eq:seq-ub-1}
I + 
\left\lceil \frac{I_1}{b} \right\rceil \cdots
\left\lceil \frac{I_N}{b} \right\rceil\cdot
R (N+1)b\text.
\end{equation}

In \cref{sec:disc-seq}, within the proof of \cref{thm:seq-attain}, we will weaken and simplify \cref{eq:seq-ub-1} for easier comparison with the lower bounds \cref{eq:seq_LB,eq:trivial-lb}.
We will assume additionally that the fast memory size $M$ is sufficiently large with respect to the tensor order $N$, but not too large with respect to the tensor dimensions $I_1,\ldots,I_N$.
Under these assumptions, picking the block size $b$ to be approximately $M^{1/N}$ gives an upper bound of the form
\begin{equation}
\label{eq:seq-ub-2}
O\left(I + \frac{NIR}{M^{1-1/N}}\right)
\text.
\end{equation}
To see how \cref{eq:seq-ub-2} might be obtained from \cref{eq:seq-ub-1}, substitute $b = (M/2)^{1/N}$, supposing $b$ is a positive integer that satisfies \cref{eq:seq-b-constraint} and divides $I_1,\ldots,I_N$.

\subsection{Parallel Stationary Tensor Algorithm}
\label{sec:par_staten}

We present two parallel algorithms, \cref{alg:par_staten,alg:par_gen}, the first of which is a special case of the second.
Here in \cref{sec:par_staten} we present the special case of \cref{alg:par_staten} in detail because its notation is simpler and we expect it to apply more frequently in typical applications, where $NR$ is small relative to $I/P$.
The general algorithm,  \cref{alg:par_gen}, is presented in \cref{sec:par_gen}.

\newcommand{\procdim}{3}
\newcommand{\proc}{\draw[black,shift={(-.5,-.5)}] (0,0) grid (\procdim,\procdim);}
\newcommand{\highlight}{gray!75}
\newcommand{\commhighlight}{gray!25}
\newcommand{\parscale}{.475}

\newcommand{\parbasepic}{
\begin{scope}[canvas is yz plane at x=.5,shift={(1.5,-1.5)}]
	\draw[fill=\highlight,shift={(0,1)}] (0,0) rectangle (1,1);
	\draw[fill=\highlight,shift={(-2.5,2-1/3)},xscale=.5] (0,0) rectangle (-1,-1/9);
	\draw[shift={(0,-1.5)},yscale=.5] (0,0) rectangle (1/9,-1);
\end{scope}
\begin{scope}[canvas is zx plane at y=(\procdim-.5),rotate=-90,shift={(-.5,-3)}]
	\draw[fill=\highlight,shift={(0,2.5)}] (0,0) rectangle (1,1);
	\draw[fill=\highlight,yscale=.5] (0,0) rectangle (1/9,-1);
\end{scope}
\begin{scope}[canvas is yx plane at z=.5,yscale=-1,rotate=0]
	\draw[fill=\highlight,shift={(1.5,-.5)}] (0,0) rectangle (1,1);
\end{scope}

\begin{scope}[canvas is yz plane at x=.5,rotate=-90]
	\proc
\end{scope}
\begin{scope}[canvas is yx plane at z=.5,yscale=-1,rotate=0]
	\proc
\end{scope}
\begin{scope}[canvas is zx plane at y=(\procdim-.5),rotate=180]
	\proc
\end{scope}

\begin{scope}[canvas is yz plane at x=.5,shift={(1.5,-.5)}]
	\draw[shift={(-2.5,0)},xscale=.5] (0,-2) grid (-1,1);
	\node[draw=none] at (-3.5,-.5) {$\Mn{A}{1}$};
	\draw[shift={(0,-2.5)},yscale=.5] (-2,0) grid (1,-1);
	\node[draw=none] at (-.5,-3.5) {$\Mn{B}{2}$};
\end{scope}
\begin{scope}[canvas is zx plane at y=(\procdim-.5),rotate=-90,shift={(1.5,-.5)}]
	\draw[shift={(0,-2.5)},yscale=.5] (-2,0) grid (1,-1);
	\node[draw=none] at (-.5,-3.5) {$\Mn{A}{3}$};
\end{scope}
}
\begin{figure*} 
\centering
  \subfloat[Start with one subtensor and subset of rows of each input matrix.]{\begin{tikzpicture}[x={(-0.5cm,-0.4cm)}, y={(1cm,0cm)}, z={(0cm,1cm)},every node/.append style={transform shape},scale=\parscale]

\parbasepic

\end{tikzpicture}} \quad
  \subfloat[All-Gather rows from $\Mn{A}{1}$ (\cref{line:par_staten:ag}).]{\begin{tikzpicture}[x={(-0.5cm,-0.4cm)}, y={(1cm,0cm)}, z={(0cm,1cm)},every node/.append style={transform shape},scale=\parscale]

\begin{scope}[canvas is yz plane at x=.5]
	\draw[fill=\commhighlight,shift={(-.5,-.5)}] (0,0) rectangle (3,1);
	\draw[fill=\commhighlight,shift={(-1,.5)},xscale=.5] (0,0) rectangle (-1,-1);
\end{scope}
 right face
\begin{scope}[canvas is zx plane at y=(\procdim-.5),rotate=-90]
	\draw[fill=\commhighlight,shift={(-.5,-.5)}] (0,0) rectangle (3,1);
\end{scope}
\begin{scope}[canvas is yx plane at z=.5,yscale=-1,rotate=0]
	\draw[fill=\commhighlight,shift={(-.5,-.5)}] (0,0) rectangle (3,3);
\end{scope}

\parbasepic

\end{tikzpicture}} \quad
  \subfloat[All-Gather rows from $\Mn{A}{3}$ (\cref{line:par_staten:ag}).]{\begin{tikzpicture}[x={(-0.5cm,-0.4cm)}, y={(1cm,0cm)}, z={(0cm,1cm)},every node/.append style={transform shape},scale=\parscale]

\begin{scope}[canvas is yz plane at x=.5]
	\draw[fill=\commhighlight,shift={(-.5,.5)}] (0,0) rectangle (3,-3);
	\draw[fill=\highlight,shift={(-1,.5)},xscale=.5] (0,0) rectangle (-1,-1);
\end{scope}
\begin{scope}[canvas is zx plane at y=(\procdim-.5),rotate=-90]
	\draw[fill=\commhighlight,shift={(-.5,.5)}] (0,0) rectangle (1,-3);
	\draw[fill=\commhighlight,yscale=.5,shift={(-.5,-6)}] (0,0) rectangle (1,-1);
\end{scope}
\begin{scope}[canvas is yx plane at z=.5,yscale=-1,rotate=0]
	\draw[fill=\commhighlight,shift={(-.5,-.5)}] (0,0) rectangle (3,1);
\end{scope}

\parbasepic

\end{tikzpicture}} \quad
  \subfloat[Compute local contribution to rows of $\Mn{B}{2}$ (\cref{line:par_staten:local}).]{\begin{tikzpicture}[x={(-0.5cm,-0.4cm)}, y={(1cm,0cm)}, z={(0cm,1cm)},every node/.append style={transform shape},scale=\parscale]


\begin{scope}[canvas is yz plane at x=.5]
	\draw[fill=\highlight,shift={(-1,.5)},xscale=.5] (0,0) rectangle (-1,-1);
\end{scope}
\begin{scope}[canvas is zx plane at y=(\procdim-.5),rotate=-90]
	\draw[fill=\highlight,yscale=.5,shift={(-.5,-6)}] (0,0) rectangle (1,-1);
\end{scope}

\begin{scope}[canvas is yz plane at x=.5]
	\draw[fill=\highlight,shift={(1.5,-3)},yscale=.5] (0,0) rectangle (1,-1);
\end{scope}

\parbasepic

\end{tikzpicture}} \quad
  \subfloat[Reduce-Scatter to compute and distribute $\Mn{B}{2}$ evenly (\cref{line:par_staten:rs}).]{\begin{tikzpicture}[x={(-0.5cm,-0.4cm)}, y={(1cm,0cm)}, z={(0cm,1cm)},every node/.append style={transform shape},scale=\parscale]


\begin{scope}[canvas is yz plane at x=.5]
	\draw[fill=\commhighlight,shift={(1.5,.5)}] (0,0) rectangle (1,-3);
	\draw[fill=\commhighlight,shift={(1.5,-3)},yscale=.5] (0,0) rectangle (1,-1);
	\draw[fill=\highlight,shift={(1.5,-3)},yscale=.5] (0,0) rectangle (1/9,-1);
	\draw[fill=\highlight,shift={(-1,.5)},xscale=.5] (0,0) rectangle (-1,-1);
\end{scope}
\begin{scope}[canvas is zx plane at y=(\procdim-.5),rotate=-90]
	\draw[fill=\commhighlight,shift={(-.5,.5)}] (0,0) rectangle (3,-3);
	\draw[fill=\highlight,yscale=.5,shift={(-.5,-6)}] (0,0) rectangle (1,-1);
\end{scope}
\begin{scope}[canvas is yx plane at z=.5,yscale=-1,rotate=0]
	\draw[fill=\commhighlight,shift={(1.5,-.5)}] (0,0) rectangle (1,3);
\end{scope}

\parbasepic

\end{tikzpicture}}
  \caption{Parallel Stationary Tensor Algorithm data distribution, communication, and computation across steps for $N=3$ and $n=1$.  Highlighted areas correspond to processor $(1,3,1)$ and its subcommunicators.}
  \label{fig:ParallelStationaryAlg} 
\end{figure*}
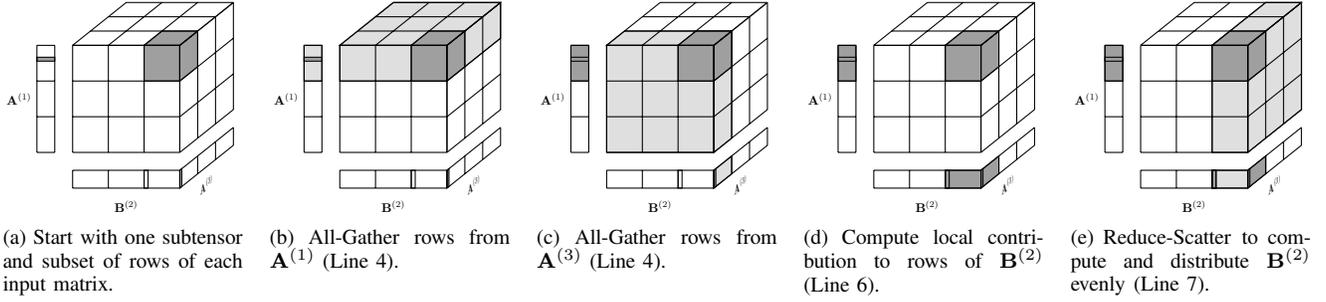

\subsubsection{Data Distribution}
\label{sec:par_staten:dist}

For an $N$-way tensor, we organize processors into an $N$-way logical processor grid. 
We factor $P = P_1P_2\cdots P_N$ and identify each processor by an $N$-tuple 
\[ \V{p} = (p_1,\ldots,p_N) \in [P_1]\times\cdots\times[P_N]\text.\]
We partition each tensor dimension $k \in [N]$ into $P_k$ parts, 
\[
[I_1] = \left\{S^{(1)}_{p_1}\right\}_{p_1 \in [P_1]}
\text,\quad
\ldots
\text,\quad
[I_N] = \left\{S^{(N)}_{p_N}\right\}_{p_N \in [P_N]}
\text.
\]
Each processor $\V{p}$ stores, 
initially (before execution),
\begin{itemize}
\item the subtensor
\[ \X_{\V{p}} = \X(S^{(1)}_{p_1},\ldots,S^{(N)}_{p_N}) \text{, and,}\]
\item for each $k \in [N] \setminus \{n\}$, a part $\Mn{A}{k}_{\V{p}}$ in a partition of   
\[ \Mn{A}{k}_{p_k} = \Mn{A}{k}(S^{(k)}_{p_k},:) \text,\]
across processors $\V{p}'$ with $p'_k=p_k$; 
\end{itemize}
during execution,
\begin{itemize}
\item the submatrices $\Mn{A}{k}_{p_k}$, $k \in [N]\setminus \{n\}$ and
\item a matrix $\M{C}_{p_n}$ the same size as (and used in the summation of) $\Mn{B}{n}_{p_n}$; and,
\end{itemize}
terminally (after execution), 
\begin{itemize}
\item a part $\Mn{B}{n}_{\V{p}}$ in a partition of
\[ \Mn{B}{n}_{p_n} = \Mn{B}{n}(S^{(n)}_{p_n}, :) \text, \]
across processors $\V{p}'$ with $p'_n=p_n$.
\end{itemize}
In words, each mode's factor matrix is distributed block-rowwise across the processor hyperslices of that mode, and each block row block is then partitioned arbitrarily across the processors in its hyperslice.
During execution, these block rows are replicated within hyperslices. 

Let us clarify a notational detail: while $\Mn{A}{k}_{p_k},\Mn{B}{n}_{p_n}$ are matrices,
the (sub)sets of matrix entries $\Mn{A}{k}_{\V{p}},\Mn{B}{n}_{\V{p}}$ need not be (sub)matrices.

\begin{algorithm}[t]
\small
\caption{Parallel Stationary MTTKRP Algorithm}
\label{alg:par_staten}
\begin{algorithmic}[1]
\Function{$\Mn{B}{n}_{\V{p}}=$ Par-Stat-MTTKRP}{$\X_{\V{p}},\{\Mn{A}{k}_{\V{p}}\},n$}
	\State $\V{p}=(p_1,\dots,p_N)$ is my processor id
	\ForAll{$k \in [N] \setminus \{n\}$}
		\State $\Mn{A}{k}_{p_k} = $ All-Gather($\Mn{A}{k}_{\V{p}},(:,\ldots,:,p_k,:,\ldots,:)$)
			\label{line:par_staten:ag}
	\EndFor
	\State $\M{C}_{p_n} = $ Local-MTTKRP($\X_{\V{p}},\{\Mn{A}{k}_{p_k}\},n$)
		\label{line:par_staten:local}
	\State $\Mn{B}{n}_{\V{p}} = $ Reduce-Scatter($\M{C}_{p_n},(:,\ldots,:,p_n,:,\ldots,:)$)
		\label{line:par_staten:rs}
\EndFunction
\end{algorithmic}
\end{algorithm}

\subsubsection{Algorithm}
\label{sec:par_staten:alg}

The pseudocode is given in \cref{alg:par_staten}.
We use the term \emph{stationary (tensor)} to describe this algorithm because the input tensor is never communicated.
Instead, each processor gathers all the input factor matrix data that participates in \mults\ involving the local tensor entries.
Then, the local computation is itself an MTTKRP.
To compute the output of the global MTTKRP, processors again must communicate to reduce values that correspond to the same output matrix entries.
The data distributions are organized using an $N$-way processor grid so that the gathers and reduction are performed across processor hyperslices using collective communication operations All-Gather and Reduce-Scatter.

\subsubsection{Analysis}
\label{sec:par_staten:analysis}

We analyze the communication cost first.
Communication occurs only in the All-Gather and Reduce-Scatter collectives in \cref{line:par_staten:ag,line:par_staten:rs}.
Each processor $\V{p}$ is involved in 
$N{-}1$ All-Gathers (\cref{line:par_staten:ag}, $k \in [N]\setminus\{n\}$) and 
one Reduce-Scatter (\cref{line:par_staten:rs}).
Over all processors, 
\cref{line:par_staten:ag} ($k \in [N]\setminus\{n\}$) specifies $P_k$ simultaneous All-Gathers, and 
\cref{line:par_staten:rs} specifies $P_n$ simultaneous Reduce-Scatters,
one for each hyperslice of the processor grid normal to the $k$-th dimension.

Towards an upper bound, we suppose the collectives are performed in a blocking manner.
For any $N{-}1$ All-Gathers in \cref{line:par_staten:ag} ($k \in [N]\setminus\{n\}$) and any Reduce-Scatter in \cref{line:par_staten:rs}, there exists a processor $\V{p}$ involved in all.
This justifies our upper bound approach, to examine any $N$ collectives each with maximal communication cost among the $P_k$ performed simultaneously (we need not specify the common processor $\V{p}$).

Since we do not quantify latency cost in this work, we will use the simpler \emph{bucket algorithms}.
A bucket All-Gather or Reduce-Scatter algorithm with $q$ processors 
proceeds in $q{-}1$ steps, at each of which each processor passes left an array of size at most $w$.
That is, $w$ is the largest local array size before (All-Gather) or after (Reduce-Scatter) the collective.  
The communication cost is at most $(q-1)w$, which is (bandwidth-) optimal for perfectly balanced data distributions \cite{CH+07}.
For Reduce-Scatter, there is also an arithmetic cost of at most $(q-1)w$ operations (here, additions).

In the present cases, we have 
$q = P/P_k$ for \cref{line:par_staten:ag} ($k \in [N] \setminus\{n\}$) and \cref{line:par_staten:rs} ($k = n$).
The local vector size $w$ depends on the data distributions specified in \cref{sec:par_staten:dist}:%
\[
w \le \begin{cases}
\max_{\V{p}} \nnz(\Mn{A}{k}_{\V{p}}) & k \in [N] \setminus\{n\} \\
\max_{\V{p}} \nnz(\Mn{B}{n}_{\V{p}}) & k = n 
\text.\end{cases}
\]
The overall communication cost is thus bounded above,
\begin{equation}
\label{eq:par_staten:comm-ub-1}
\sum_{k \in [N]} 
\left(\frac{P}{P_k}-1\right)
\cdot
\begin{cases}
\displaystyle\max_{\V{p}} \nnz(\Mn{A}{k}_{\V{p}}) & k \ne n \\
\displaystyle\max_{\V{p}} \nnz(\Mn{B}{n}_{\V{p}}) & k = n 
\text.\end{cases}
\end{equation}

The arithmetic cost is bounded above in terms of the costliest local MTTKRP (\cref{line:par_staten:local}) and the costliest Reduce-Scatter (\cref{line:par_staten:rs}): the number of operations is at most
\begin{equation}
\label{eq:par_staten:flops-ub-1}
NR \max_{\V{p}} \left(\prod_{k \in [N]} |S^{(k)}_{p_k}|\right)
+
\left(\frac{P}{P_n}-1\right)
\max_{\V{p}}\left( \nnz(\Mn{B}{n}_{\V{p}})\right)
\text.
\end{equation}

The per-processor storage cost is bounded above,
\begin{equation}
\label{eq:par_staten:mem-ub-1}
 \max_{\V{p}} 
\left(
\prod_{k \in [N]} |S^{(k)}_{p_k}|
+
\sum_{k \in [N]} |S^{(k)}_{p_k}|R 
\right)
\text.
\end{equation}

Assuming we can choose a processor grid such that $P_k \approx I_k / (I/P)^{1/N}$ and divides $I_k$ evenly, we choose the data distribution such that $|S^{(k)}_{p_k}| = I_k/P_k$ for $k\in[N]$, which simplifies these upper bounds:
the communication cost bound \cref{eq:par_staten:comm-ub-1} is
\begin{equation*}
O\left(NR\left(\frac{I}{P}\right)^{1/N}\right)\text,
\end{equation*}
the arithmetic cost bound \cref{eq:par_staten:flops-ub-1} is
\begin{equation*}
O\left(\frac{NIR}{P}\right)\text,
\end{equation*}
and the (per-processor) storage cost bound \cref{eq:par_staten:mem-ub-1} is 
\begin{equation*}
O\left(\frac{I}{P} + NR\left(\frac IP\right)^{1/N}\right)\text.
\end{equation*}
We weaken these assumptions on the processor grid and make them more explicit in the proof of \cref{thm:par-attain}.

We note that to save some arithmetic, the algorithm could break the atomicity of the \mults\ without changing the communication costs of the algorithm: each processor could precompute the explicit local Khatri-Rao product and perform a local matrix multiplication, reducing the first term in \cref{eq:par_staten:flops-ub-1} to 
\begin{equation}
\label{eq:par_staten:reducing-flops}
R \max_{\V{p}} \left(\left(\prod_{k \in [N]} |S^{(k)}_{p_k}|\right)\left(2+\frac{1}{|S^{(n)}_{p_n}|}\right)\right)
\text,
\end{equation}
which is $O(IR/P)$ with a load-balanced tensor distribution.

\subsection{Parallel General Algorithm}
\label{sec:par_gen}

This section studies \cref{alg:par_gen}, a generalization of the stationary tensor algorithm, \cref{alg:par_staten},  described in \cref{sec:par_staten}.
\Cref{alg:par_gen} parallelizes over all $N{+}1$ dimensions of the iteration space: the $N$ tensor dimensions, bounded by $I_1,\ldots,I_N$, and the matrix column dimension, bounded by $R$.
In contrast, recall that \cref{alg:par_staten} parallelizes over just the $N$ tensor dimensions.
Roughly speaking, \cref{alg:par_gen} is more communication efficient than \cref{alg:par_staten} when $NR$ is large relative to $I/P$.

\begin{algorithm}
\small
\caption{Parallel General MTTKRP Algorithm}
\label{alg:par_gen}
\begin{algorithmic}[1]
\Function{$\Mn{B}{n}_{\V{p}}=$ Par-Gen-MTTKRP}{$\X_{\V{p}},\{\Mn{A}{k}_{\V{p}}\},n$}
	\State $\V{p}=(p_0,p_1,{\dots},p_N)$ is my processor id
	\State $\X_{p_1,\ldots,p_N}$ = All-Gather($\X_{\V{p}}$,$(:,p_1,\ldots,p_N)$)
		\label{line:par_gen:agX}
	\ForAll{$k \in [N] \setminus \{n\}$}
		\State $\Mn{A}{k}_{p_k,p_0} = $ All-Gather($\Mn{A}{k}_{\V{p}}$,$(p_0,:,\ldots,:,p_k,:,\ldots,:)$)
			\label{line:par_gen:ag}
	\EndFor
	\State $\M{C}_{p_n,p_0} = $ Local-MTTKRP($\X_{p_1,\ldots,p_N}$, $\{\Mn{A}{k}_{p_k,p_0}\}$, $n$)
		\label{line:par_gen:local}
	\State $\Mn{B}{n}_{\V{p}} = $ Reduce-Scatter($\M{C}_{p_n,p_0}$,$(p_0,:,\ldots,:,p_n,:,\ldots,:)$)
		\label{line:par_gen:rs}
\EndFunction
\end{algorithmic}
\end{algorithm}

\subsubsection{Data Distribution}
\label{sec:par_gen:dist}

For an $N$-way tensor, we organize processors into an $(N{+}1)$-way logical processor grid. 
We factor $P = P_0P_1P_2\cdots P_N$ and identify each processor by an ($N{+}1$)-tuple 
\[ \V{p} = (p_0,p_1,\ldots,p_N) \in [P_0] \times [P_1]\times\cdots\times[P_N]\text.\]
As before, we partition each tensor dimension $k \in [N]$ into $P_k$ parts, 
\[
[I_1] = \left\{S^{(1)}_{p_1}\right\}_{p_1 \in [P_1]}
\text,\quad
\ldots
\text,\quad
[I_N] = \left\{S^{(N)}_{p_N}\right\}_{p_N \in [P_N]}
\text.
\]
Additionally we now partition the matrix column dimension into $P_0$ parts,
\[ [R] = \{T_{p_0}\}_{p_0 \in [P_0]} \text. \]
Each processor $\V{p}$ stores, initially (before execution),
\begin{itemize}
\item a part $\X_{\V{p}}$ in a partition of
\[ \X_{p_1,\ldots,p_N} = \X(S^{(1)}_{p_1},\ldots,S^{(N)}_{p_N})\text, \]
across processors $\V{p}'$ with $p_k'=p_k$ ($k \in [N]$), and
\item for each $k \in [N] \setminus \{n\}$, a part $\Mn{A}{k}_{\V{p}}$ in a partition of  
\[ \Mn{A}{k}_{p_k,p_0} = \Mn{A}{k}(S^{(k)}_{p_k},T_{p_0}) \text,\]
across processors $\V{p}'$ with $p'_0=p_0$ and $p'_k=p_k$; 
\end{itemize}
during execution,
\begin{itemize}
\item the subtensor $\X_{p_1,\ldots,p_N}$, 
\item the submatrices $\Mn{A}{k}_{p_k,p_0}$, $k \in [N]\setminus \{n\}$, and $\Mn{B}{n}_{p_n,p_0}$, and
\item a matrix $\M{C}_{p_n,p_0}$ the same size as (and used in the summation of) $\Mn{B}{n}_{p_n,p_0}$; and,
\end{itemize}
terminally (after execution),
\begin{itemize}
\item a part $\Mn{B}{n}_{\V{p}}$ in a partition of  
\[ \Mn{B}{n}_{p_n,p_0} = \Mn{B}{n}(S^{(n)}_{p_n}, T_{p_0}) \text,\]
across processors $\V{p}'$ with $p'_0=p_0$ and $p_n'=p_n$.
\end{itemize}

Let us clarify a notational detail: while $\X_{p_1,\ldots,p_N},\Mn{A}{k}_{p_k},\Mn{B}{n}_{p_n}$ are tensors/matrices,
the (sub)sets of tensor/matrix entries $\X_{\V{p}},\Mn{A}{k}_{\V{p}},\Mn{B}{n}_{\V{p}}$ need not be (sub)tensors/matrices.

\subsubsection{Algorithm}
\label{sec:par_gen:alg}

As mentioned at the beginning of \cref{sec:par_staten}, the general algorithm \cref{alg:par_gen} parallelizes over all $N{+}1$ dimensions of the iteration space: unlike the stationary algorithm \cref{alg:par_staten}, entries of the tensor $\X$ are now communicated among processors.
One can think of \cref{alg:par_gen} as logically dividing the output factor matrix $\Mn{B}{n}$ into $P_0$ block-columns, each assigned to a separate subset of $P/P_0$ processors, and running \cref{alg:par_staten} on each subset of processors.

The structure of \cref{alg:par_gen} is very similar to \cref{alg:par_staten}: each processor gathers the necessary input data, performs local computation, and then participates in a Scatter-Reduce to compute and redistribute the output.
The main difference is in \cref{line:par_gen:agX}, where a subtensor is All-Gathered across the $P_0$ processors in each processor grid fiber along the last grid dimension.

\subsubsection{Analysis}
\label{sec:par_gen:analysis}

The analysis of \cref{alg:par_gen} is nearly identical to that of \cref{alg:par_staten}; see \cref{sec:par_staten:analysis}.

The overall communication cost is bounded above,
\begin{multline}
\label{eq:par_gen:comm-ub-1}
\left(P_0 - 1\right) \cdot \max_{\V{p}} \nnz(\X_{\V{p}}) 
\\+
\sum_{k \in [N]} 
\left(\frac{P}{P_0P_k}-1\right)
\cdot
\begin{cases}
\displaystyle\max_{\V{p}} \nnz(\Mn{A}{k}_{\V{p}}) & k \ne n \\
\displaystyle\max_{\V{p}} \nnz(\Mn{B}{n}_{\V{p}}) & k = n 
\text.\end{cases}
\end{multline}
Comparing with \cref{eq:par_staten:comm-ub-1}, we notice the (new) first term, due to the new All-Gather (\cref{line:par_gen:agX}), as well as the modified costs of the other collectives, which are now performed on lower dimensional processor grid hyperslices. 

The arithmetic cost is bounded above by
\begin{multline}
\label{eq:par_gen:flops-ub-1}
N \max_{\V{p}} \left(|T_{p_0}| \cdot \prod_{k \in [N]} |S^{(k)}_{p_k}|\right)
\\+
\left(\frac{P}{P_0P_n}-1\right)
\max_{\V{p}} \nnz(\Mn{B}{n}_{\V{p}})
\text.
\end{multline}
Comparing with \cref{eq:par_staten:flops-ub-1}, we notice that the first term has changed due to blocking in the matrix column dimension, and the second term has changed for the same reason as the communication cost.

The per-processor storage cost is bounded above by
\begin{equation}
\label{eq:par_gen:mem-ub-1}
\max_{\V{p}} 
\left(\prod_{k \in [N]} |S^{(k)}_{p_k}|
+
\sum_{k \in [N]} |S^{(k)}_{p_k}|\cdot|T_{p_0}|  
\right) 
\text.
\end{equation}
Comparing with \cref{eq:par_staten:mem-ub-1}, we notice that the second term has changed due to distributing matrix columns.

Assuming we can choose a processor grid such that $P_0\approx (NR)^{N/(2N-1)}/(I/P)^{(N-1)/(2N-1)}$ and $P_k \approx I_k / (IP_0/P)^{1/N}$ for $k\in [N]$, and that we can choose the data distribution such that $|S^{(k)}_{p_k}| = I_k/P_k$, $|T_{p_0}| = R/P_0$, $\nnz(\X_{\V{p}}) = I/P$, $\nnz\Mn{A}{k}_{\V{p}} = I_kR/P$, and $\nnz\Mn{B}{n}_{\V{p}} = I_nR/P$ (assuming everything divides evenly), these upper bounds can be simplified.
The communication cost bound \cref{eq:par_gen:comm-ub-1} and the storage cost bound \cref{eq:par_gen:mem-ub-1} are
\begin{equation*}
O\left(NR\left(\frac IP\right)^{1/N}+\left(\frac{NIR}{P}\right)^{\frac{N}{2N-1}}\right)\text,
\end{equation*}
and the arithmetic cost bound \cref{eq:par_gen:flops-ub-1} is
\begin{equation*}
O\left(\frac{NIR}{P}\right)\text.
\end{equation*}

The trick for reducing arithmetic discussed in \cref{sec:par_staten:analysis}, breaking atomicity of the \mults, applies here as well.
The result is reducing the first term in the upper bound \cref{eq:par_gen:flops-ub-1} to
\begin{equation*}
\max_{\V{p}} \left( \left(|T_{p_0}|\cdot \prod_{k \in [N]} |S^{(k)}_{p_k}|\right)\left(2+\frac{1}{|S^{(n)}_{p_n}|}\right)\right) \text,
\end{equation*}
which is $O(IR/P)$ assuming a load-balanced distribution.

\section{Discussion}
\label{sec:disc}

\subsection{Sequential Case}
\label{sec:disc-seq}

We would like to compare the upper bound,
\begin{equation}
\label{eq:Wub}
W_{ub} = I + (N+1)\left(\prod_{k \in [N]} \left\lceil\frac{I_k}{b}\right\rceil\right)b R 
\text,
\end{equation}
valid for any $b\in\{1,2,\ldots\}$ satisfying
\begin{equation}
\label{eq:b-constraint}
M \ge b^N + Nb
\text,
\end{equation}
with the memory dependent lower bound (\cref{thm:seq_LB})
\begin{equation}
\label{eq:Wlb1} 
W_{lb1}=\frac{NIR}{3(3M)^{1-1/N}}-M
\text,
\end{equation}
and the trivial lower bound (\cref{fct:trivial-lb})
\begin{equation}
\label{eq:Wlb2} 
W_{lb2}=I+\sum_{k\in[N]}I_kR-2M
\text.
\end{equation}

We now show that under certain assumptions on $M$, for example assuming that the tensor is too large to fit in fast memory, the upper bound and lower bounds differ by no more than a constant.

\begin{theorem}
\label{thm:seq-attain}
Suppose $M$ is sufficiently larger than the number of dimensions $N$ and that each dimension $I_k$ is sufficiently larger than $M^{1/N}$.  Then \cref{alg:seq-blocked} is communication optimal to within a constant factor.
\end{theorem}
\begin{IEEEproof}
Suppose there exist positive constants $\alpha,\beta,\gamma,\delta,\epsilon$ such that
{\footnotesize\begin{align}
M &\ge \left(\frac{N\alpha^{1/N}}{1-\alpha}\right)^{\frac{N}{N-1}}
&
\alpha &< 1 
\label{eq:hyp1}\\
M &\ge \left(\frac{1}{\alpha^{1/N}-\beta^{1/(N-1)}}\right)^N
&
\beta &< \alpha^{1-1/N}
\label{eq:hyp2}\\
M &\le \left(\frac{\left(\frac{N}{N+1}\gamma\right)^{1/N}-1}{\alpha^{1/N}}\min_{k \in [N]} I_k\right)^N
&
\gamma &> 1+\frac1N
\label{eq:hyp3}\\
M &\le \frac12\left((1-\delta)I + \sum_{k \in [N]} I_k R\right)
&
\delta &< 1+\sum_{k \in [N]} \frac{I_k}{I}R  
\label{eq:hyp4}\\
M &\le \left(\left(\frac{1}{3^{2-1/N}}-\epsilon\right)NIR\right)^{\frac{N}{2N-1}} 
&
\epsilon &< \frac{1}{3^{2-1/N}}
\text.
\label{eq:hyp5}
\end{align}}
For \cref{alg:seq-blocked}, we choose block size 
\[ b = \left\lfloor \left( \alpha M \right)^{1/N} \right\rfloor. \]
It follows from \cref{eq:hyp1} that $b$ satisfies \cref{eq:b-constraint}.
It follows from \cref{eq:hyp2} that $b \ge 1$ and, moreover, 
\[ b^{N-1} \ge \beta M^{1-1/N}\text.\]
It follows from \cref{eq:hyp3} that 
\[ \prod_{k \in [N]}\left\lceil\frac{I_k}{b}\right\rceil \le \gamma\frac{I}{b^N}\frac{N}{N+1}\text.\]
Since $\beta < 1 < \gamma$, it then follows that
\[ W_{ub} \le \frac{\gamma}{\beta}\left(I + \frac{NIR}{M^{1-1/N}}\right) \text.\]

It follows from \cref{eq:hyp4} that 
\[ W_{lb2} \ge \delta I \text.\]
It follows from \cref{eq:hyp5} that
\[ W_{lb1} \ge \epsilon \frac{NIR}{M^{1-1/N}} \text.\]
Since these are positive lower bounds,
\[ \max(W_{lb1},W_{lb2}) \ge \frac{\min(\delta,\epsilon)}{2}\left(I  + \frac{NIR}{M^{1-1/N}} \right) > 0\text,\]
which matches the upper bound to within a constant factor.
%
\end{IEEEproof}

To illustrate the hypotheses \cref{eq:hyp1,eq:hyp2,eq:hyp3,eq:hyp4,eq:hyp5} of \cref{thm:seq-attain}, take, for example, the constants $\beta = 1-\alpha = 1/100$,  $\gamma=100$, and $\delta=\epsilon=1/10$, which satisfy the right-hand inequalities for all fast memory sizes $M$ and problem parameters $N,I_1,\ldots,I_N,R$.
Clearly there are infinitely many choices of $M$ and the problem parameters that satisfy the left-hand inequalities.
For example, supposing $N \le 10$ and $I_1 = I_2 = \cdots = I_N$, 
the left-hand inequalities require that the fast memory size $M$ is 
bounded below by $10^4$ (due to \cref{eq:hyp1,eq:hyp2}), and 
above by the minimum of $I/1000$ (due to \cref{eq:hyp3,eq:hyp4}) and $\sqrt{NIR}/10$ (due to \cref{eq:hyp5}).
We claim that this example includes parameters that are representative of real-world machines and problems of practical interest.
Of course, since we have placed a constant upper bound on $N$, this example does not illustrate (asymptotic) behavior with respect to $N$.

We also compare the communication cost of \cref{alg:seq-blocked}, $O(I+NIR/M^{1-1/N})$, with the MTTKRP via matrix multiplication approach.
We assume a communication-optimal matrix multiplication is used, achieving $O(I+IR/M^{1/2})$ communication cost and performing $2IR$ operations.
Here, the cost of explicitly forming the Khatri-Rao product matrix is a lower order term, assuming $R<I_k$ for all $k\in [N]$.
Assuming $N=O(M^{1/2-1/N})$, the communication cost of \cref{alg:seq-blocked} never exceeds that of MTTKRP via matrix multiplication.

If the communication cost is dominated by accessing the tensor elements (i.e., $R=O(M^{1/2})$), then both approaches perform the same amount of communication and \cref{alg:seq-blocked} performs a factor of $N/2$ more computation.
If the communication cost is dominated by repeatedly accessing the factor matrix elements (i.e., $NR=\Omega(M^{1-1/N})$), then \cref{alg:seq-blocked} is more efficient, requiring a factor of $O(M^{1/2-1/N}/N)$ less communication.

In practice, we expect $N$ to be very small relative to $M$, so the assumption $N=O(M^{1/2-1/N})$ is mild.
However, we also expect $R$ to be small relative to $M$, and in that case, the dominant communication cost of reading tensor elements from memory is shared by both approaches.
In this case, the matrix multiplication approach benefits from fewer operations, and in practice it can also exploit highly tuned software for matrix multiplication.

\subsection{Parallel Case}
\label{sec:disc-par}

Recall from \cref{sec:par_staten,sec:par_gen} that we presented two parallel algorithms, \cref{alg:par_staten,alg:par_gen}, the former being the special case of the latter with $P_0=1$.

The communication upper bound for \cref{alg:par_gen}, 
\begin{multline}
\label{eq:par-Wub}
\left(P_0 - 1\right) \cdot \max_{\V{p}} \nnz(\X_{\V{p}}) 
\\+
\sum_{k \in [N]} 
\left(\frac{P}{P_0P_k}-1\right)
\cdot
\begin{cases}
\displaystyle\max_{\V{p}} \nnz(\Mn{A}{k}_{\V{p}}) & k \ne n \\
\displaystyle\max_{\V{p}} \nnz(\Mn{B}{n}_{\V{p}}) & k = n 
\text.\end{cases}
\end{multline}
is valid for any factorization $P=P_0P_1\cdots P_N$ and data distribution specified in \cref{sec:par_gen:dist}.
(Recall that $\X_{\V{p}}$, $\Mn{A}{k}_{\V{p}}$ ($k \in [N] \setminus \{n\}$), and $\Mn{B}{n}_{\V{p}}$ denote the distributed subsets of tensor and factor matrix entries.)
We wish to compare this upper bound with the lower bound from \cref{thm:MemIndepLB} 
\begin{equation}
\label{eq:par-Wlb1}
2\left(\frac{NIR}{P}\right)^{\frac{N}{2N-1}}-\gamma\frac{I}{P}-\delta\sum_{k\in[N]}\frac{I_kR}{P}\text,
\end{equation}
and the lower bound from \cref{thm:rectMemIndepLB},
\begin{equation}
\label{eq:par-Wlb2}
\min\left\{\sqrt{\frac{2}{3\gamma}}NR\left(\frac{I}{P}\right)^{1/N}-\delta \sum_{k \in [N]} \frac{I_kR}{P},\frac{\gamma I}{2P}\right\}\text.
\end{equation}

\begin{theorem}
\label{thm:par-attain}
Suppose the number of processors $P$ is sufficiently large and factorable, and suppose that the tensor dimensions and rank $R$ are sufficiently large with respect to $P$.  
Then \cref{alg:par_gen} is communication optimal to within a constant factor.
\end{theorem}
\begin{IEEEproof}
To instantiate $W_\text{ub}^\text{par}$, we must specify a processor grid (i.e., a factorization of $P$ into a product $P_0P_1\cdots P_N$ of positive integers) as well as the distributions of the tensor and factor matrices.
For any processor grid, recalling the notation of \cref{sec:par_gen:dist}, we can define a data distribution where, for each processor $\V{p}$,
\begin{equation}
\label{eq:par-dist-hyp-NK}
\begin{aligned}
\nnz(\X_{\V{p}}) &\le \big\lceil \prod_k \lceil I_k/P_k \rceil/P_0 \big\rceil \text,\\
\nnz(\Mn{A}{k}_{\V{p}}) &\le \big\lceil\lceil I_k/P_k\rceil\lceil R/P_0\rceil/(P/(P_kP_0))\big\rceil \text,\\
\nnz(\Mn{B}{n}_{\V{p}}) &\le \big\lceil\lceil I_n/P_n\rceil\lceil R/P_0\rceil/(P/(P_nP_0))\big\rceil\text.
\end{aligned}
\end{equation}

To instantiate $W_\text{lb1}^\text{par},W_\text{lb2}^\text{par}$, we must assume that that no processor owns more than $\gamma I/P$ tensor entries or $\delta \sum_k I_kR/P$ factor matrix entries, for some constants $\gamma,\delta \ge 1$.
For any $\gamma,\delta > 1$, we can manipulate the upper bounds in \cref{eq:par-dist-hyp-NK} to derive relations on the machine and problem parameters such that these balance constraints hold.
In particular, we suppose there exist constants $\alpha,\beta > 1$ such that $\gamma > \alpha$, $\delta > \alpha^{1/N}\beta$, and, for all $k \in [N]$,
\begin{equation}
\label{eq:par-additional-hyp}
\begin{aligned}
P_k &\le (\alpha^{1/N}-1)I_k\text,
&
P   &\le (\gamma - \alpha) I\text,
\\
P_0 &\le (\beta-1)R\text,
&
P   &\le (\delta-\alpha^{1/N}\beta)I_kR\text.
\end{aligned}
\end{equation}
These hypotheses also yield a simpler upper bound,
\begin{equation}
\label{eq:par-Wubsimple}
W_\text{ub}^\text{par}
\le 
\gamma (P_0-1)\frac{I}{P} + \delta \sum_{k\in[N]} \frac{I_kR}{P}
\text.
\end{equation}

We now consider two cases, when $NR \le (I/P)^{1-1/N}$ and when $NR > (I/P)^{1-1/N}$.
In each case, under additional hypotheses, \cref{eq:par-Wubsimple} attains one of the two lower bounds \cref{eq:rectMemIndepLB,eq:MemIndepLB}.

In the first case, $NR \le (I/P)^{1-1/N}$, we suppose there exists a constant $\epsilon > 0$ such that $P$ factors as $P_0P_1\cdots P_N$ with $P_0=1$ and, for all $k \in [N]$, $I_k/P_k \le (\epsilon/\delta) (I/P)^{1/N}$.
Additionally, we suppose there exists a constant $\eta$, $0 < \eta < \sqrt{2/(3\gamma)}$ such that
\[ P \ge \left(\frac{\delta}{\sqrt{2/(3\gamma)}-\eta} \frac{\sum I_k}{NI^{1/N}}\right)^{\frac{N}{N-1}} \text.\]
The first hypothesis simplifies the upper bound \cref{eq:par-Wubsimple} to
$
W_\text{ub}^\text{par} \le \epsilon\cdot NR(I/P)^{1/N}\text,
$
while the second hypothesis simplifies the lower bound \cref{eq:rectMemIndepLB} to 
$
W_\text{lb2}^\text{par} \ge \eta \cdot NR(I/P)^{1/N}.
$

In the second case, $(NR)^N > (I/P)^{N-1}$, we suppose there exist constants $\mu,\nu > 0$ such that $P$ factors as $P_0P_1\cdots P_N$,
\[ 
\frac{\delta}{\nu} \left(\frac{(NR)^{N-1}}{(I/P)^N}\right)^{\frac{1}{2N-1}}\frac{I_k}{P_k} 
\le P_0 \le 
\frac{\mu}{\gamma} \left(\frac{(NR)^N}{(I/P)^{N-1}}\right)^{\frac{1}{2N-1}}\text, 
\]
for each $k \in [N]$. 
Additionally, we suppose there exists a constant $\tau$, $0 < \tau < 2-\gamma$, such that
\[ P \ge \frac{\left(\frac{\delta}{2-(\gamma+\tau)}\sum I_k\right)^{\frac{2N-1}{N-1}}R}{(NI)^{\frac{N}{N-1}}} \text.\]
The first hypothesis simplifies the upper bound \cref{eq:par-Wubsimple} to
$W_\text{ub}^\text{par} \le (\mu + \nu) \cdot (NIR/P)^{N/(2N-1)}$,
while the second hypothesis simplifies the lower bound \cref{eq:MemIndepLB} to
$W_\text{lb1}^\text{par} \ge \tau \cdot (NIR/P)^{N/(2N-1)}$.
In each of the two cases, the gap is a constant factor.
\end{IEEEproof}

To illustrate the hypotheses of \cref{thm:par-attain}, we set $\gamma=\delta=1.75$, $\alpha^{1/N}=1.05$, and $\beta=1.5$ and assume $3\leq N\leq 10$, for example, and the assumptions in \cref{eq:par-additional-hyp} for the upper bound simplification to apply become $P_k \leq 0.05I_k$, $P \leq 0.7I$, $P_0 \leq 0.5R$, and $P \leq 0.175 I_kR$.
With $\eta=\tau=0.1$ and assuming $I_k=I^{1/N}$ for all $k$, the assumptions necessary for the lower bound simplifications to apply become $P\geq 7$ and $P\geq 465NR/I^{1-1/N}$, respectively.

We also compare \cref{alg:par_gen} with the MTTKRP via matrix multiplication approach.
For comparison, we use the theoretical costs of communication-optimal parallel matrix multiplication algorithms \cite{DE+13}. 
We assume the Khatri-Rao product matrix is constructed explicitly without communication and in the distribution required to achieve the optimal communication costs of the matrix multiplication.
For simplicity, we consider the case that $I_k=I^{1/N}$ for all $k \in [N]$.
As in the case of our parallel algorithm, the optimal choice of matrix multiplication algorithm depends on the relative size of $P$, yielding many cases for comparison.

\begin{figure}
\centering
\begin{tikzpicture}
\newcommand{\datafile}{modeled.dat}
\begin{axis}[
	title=Modeled Strong-Scaling Comparison,
	xlabel=Processors,
	ylabel=Words Communicated,
	width=\columnwidth,
	xmode=log,
	log basis x={2},
	xmin=1,xmax=2^31,
	xtick={1,32,1024,32768,1048576,33554432,1073741824},
	ymode=log,
	legend pos=south west
	]
	\addplot [black,thick,dashed,mark options={solid},mark=x] table [x={P}, y={MM}] {\datafile};
	\addlegendentry{Matrix Multiplication}
	\addplot [red,thick,dashed,mark options={solid},mark=asterisk] table [x={P}, y={Alg3}] {\datafile};
	\addlegendentry{Stationary (\cref{alg:par_staten})}
	\addplot [blue,thick,dashed,mark options={solid},mark=o] table [x={P}, y={Alg4}] {\datafile};
	\addlegendentry{General (\cref{alg:par_gen})}
\end{axis}
\end{tikzpicture}
\caption{Model of strong-scaling communication performance comparing \cref{alg:par_staten}, \cref{alg:par_gen}, and MTTKRP via matrix multiplication for a 3-way cubical tensor where $I = 2^{45}$ and $R = 2^{15}$.  
The matrix multiplication costs are computed using the CARMA algorithm \cite{DE+13}, but they do not include the communication costs of forming the Khatri-Rao product.
The maximum number of processors considered is set to be the number of elements in a factor matrix.}
\label{fig:modeled}
\end{figure}
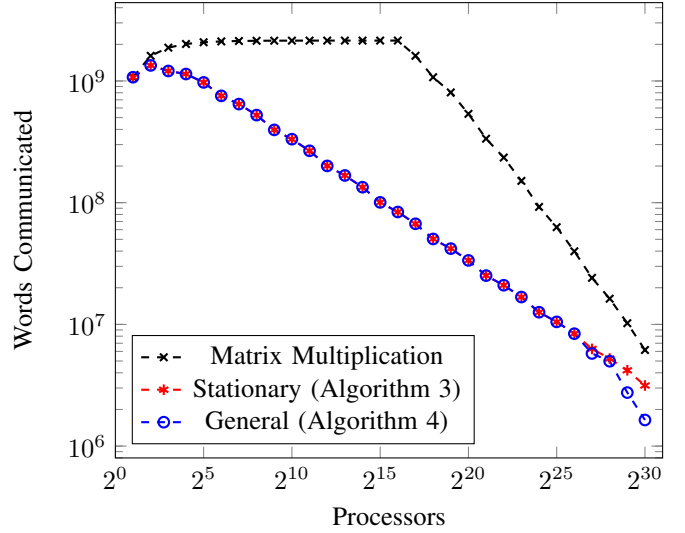

We consider only the extreme cases, ``small $P$'' and ``large $P$'', though we expect our algorithm to yield benefits in all cases.
For parallel multiplication of matrices of dimensions $I^{1/N}\times I^{N-1}$ and $I^{N-1}\times R$, if $P\leq I^{1-1/N}$, then the communication cost is $I^{1/N}R$, and if $P\geq I/R^2$, then the communication cost is $(IR/P)^{2/3}$, assuming enough memory is available \cite{DE+13}.
For comparison, if $P\leq I/(NR)^{N/(N-1)}$, then \cref{alg:par_gen} (which reduces to \cref{alg:par_staten} in this case) is optimal with communication cost $NR(I/P)^{1/N}$; if $P\geq I/(NR)^{N/(N-1)}$, then \cref{alg:par_gen} is optimal with communication cost $(NIR/P)^{N/(2N-1)}$.

Thus, we define the small $P$ case by $P \leq \min\left(I^{1-1/N},\,I/(NR)^{N/(N-1)}\right)$ and the large $P$ case by $P \geq \max\left(I/R^2,\,I/(NR)^{N/(N-1)}\right)$.
In the small $P$ case, our algorithm performs a factor of $O(P^{1/N}/N)$ less communication than MTTKRP via matrix multiplication.
In the large $P$ case, our algorithm performs a factor of $O((IR/P)^{(N-2)/(6N-3)}/N^{N/(2N-1)})$ less communication.
Again, this comparison ignores the communication cost required to form the explicit Khatri-Rao product assuming only one copy of the input matrices are initially distributed across processors.

\Cref{fig:modeled} provides a concrete comparison for a particular case, where $I_1{=}I_2{=}I_3{=}R{=}2^{15}$ and the number of processors ranges from $2^0$ up to $2^{30}$.
We see that our proposed algorithms perform less communication than matrix multiplication throughout the range of processors, and that \cref{alg:par_staten} and \cref{alg:par_gen} diverge only when $P\geq 2^{27}$.
When there are $2^{17}{=}131{,}072$ processors, \cref{alg:par_staten} and \cref{alg:par_gen} perform approximately $25\times$ less communication than the matrix multiplication approach.
This illustrates the benefits of exploiting the multi-way structure of the computation and the observation that \cref{alg:par_staten} is sufficient for most practical problems.
We note that the kink in the matrix multiplication curve is due to a switch from a 1D parallel algorithm (``1 large dimension" case) to a 2D parallel algorithm (``2 large dimension" case) and that these communication costs are optimal for matrix multiplication, up to constant factors \cite{DE+13}.
We also note that for $P>2^{30}$, which is the number of elements in each factor matrix, the All-Gather and Reduce-Scatter collectives require more efficient algorithms than the ones described in \cref{sec:algs}.

In summary, the main disadvantage of the matrix multiplication approach is that the Khatri-Rao product is treated as a general matrix despite the fact that its structure means that it depends on fewer parameters and therefore can be communicated more efficiently (in fewer words) across processors.

\section{Conclusion}
\label{sec:conc}

Because efficient algorithms and high performance implementations exist for matrix computations, it is reasonable to recast tensor computations as matrix computations.
However, the lower bounds proved in this work demonstrate an opportunity to avoid communication by exploiting the structure of the tensor computation itself.
In particular, we have shown how to extend a lower bound approach for generic programs \cite{CDKSY13} for a particular tensor computation known as MTTKRP, which is the bottleneck for algorithms that compute CP decompositions.
By demonstrating (optimal) algorithms that attain these lower bounds, we have identified a design space for implementations that we expect to achieve high performance in practice.

In many applications, the rank $R$ is small relative to the tensor dimensions.
When $R$ is also small relative to the fast memory size $M$, as discussed in \cref{sec:disc-seq}, we expect only limited practical benefits of the sequential algorithm (\cref{alg:seq-blocked}). 
However, we believe the parallel algorithms will be very competitive in practice.
The simpler algorithm (\cref{alg:par_staten}) may be the most useful, particularly when $R$ is small.
However, the general algorithm (\cref{alg:par_gen}) will likely perform better for large numbers of processors, even when $R$ is small.
The parallel data distributions are also natural ones for tensors, generalizing distributions already used for other parallel tensor computations \cite{ABK16}.

While this work focuses on a single MTTKRP computation (corresponding to a single mode), the computation nearly always occurs in the context of an optimization algorithm that requires repeatedly computing MTTKRP for each mode of the tensor.
In this context, it is beneficial to optimize across multiple MTTKRP computations, because they share both data and intermediate computations \cite{PTC13}.
Thus, optimizing over multiple MTTKRPs can save both communication and computation.

Our communication lower-bound approach extends to algorithms for multiple MTTKRPs.
Extensions are possible for other related computational kernels, such as those within algorithms for computing Tucker and other decompositions.
Another natural extension is MTTKRPs involving sparse tensors: in this case, the communication requirements depend on the nonzero structure and can be expressed in terms of a hypergraph partitioning problem \cite{BDKS16,KU15}.

\section*{Acknowledgment}

This work has been funded in part by the Laboratory-Directed Research \& Development (LDRD) program at Sandia National Laboratories. Sandia National Laboratories is a multimission laboratory managed and operated by National Technology and Engineering Solutions of Sandia, LLC., a wholly owned subsidiary of Honeywell International, Inc., for the U.S.\ Department of Energy's National Nuclear Security Administration under contract DE-NA-0003525.

\IEEEtriggeratref{17}

\bibliographystyle{IEEEtran}

\begin{thebibliography}{10}
\providecommand{\url}[1]{#1}
\csname url@samestyle\endcsname
\providecommand{\newblock}{\relax}
\providecommand{\bibinfo}[2]{#2}
\providecommand{\BIBentrySTDinterwordspacing}{\spaceskip=0pt\relax}
\providecommand{\BIBentryALTinterwordstretchfactor}{4}
\providecommand{\BIBentryALTinterwordspacing}{\spaceskip=\fontdimen2\font plus
\BIBentryALTinterwordstretchfactor\fontdimen3\font minus
  \fontdimen4\font\relax}
\providecommand{\BIBforeignlanguage}[2]{{%
\expandafter\ifx\csname l@#1\endcsname\relax
\typeout{** WARNING: IEEEtran.bst: No hyphenation pattern has been}%
\typeout{** loaded for the language `#1'. Using the pattern for}%
\typeout{** the default language instead.}%
\else
\language=\csname l@#1\endcsname
\fi
#2}}
\providecommand{\BIBdecl}{\relax}
\BIBdecl

\bibitem{KB09}
\BIBentryALTinterwordspacing
T.~G. Kolda and B.~W. Bader, ``Tensor decompositions and applications,''
  \emph{SIAM Review}, vol.~51, no.~3, pp. 455--500, September 2009. [Online].
  Available: \url{http://epubs.siam.org/doi/abs/10.1137/07070111X}
\BIBentrySTDinterwordspacing

\bibitem{AGHKT14}
\BIBentryALTinterwordspacing
A.~Anandkumar, R.~Ge, D.~Hsu, S.~M. Kakade, and M.~Telgarsky, ``Tensor
  decompositions for learning latent variable models,'' \emph{Journal of
  Machine Learning Research}, vol.~15, pp. 2773--2832, 2014. [Online].
  Available: \url{http://jmlr.org/papers/v15/anandkumar14b.html}
\BIBentrySTDinterwordspacing

\bibitem{SL+17}
N.~D. Sidiropoulos, L.~D. Lathauwer, X.~Fu, K.~Huang, E.~E. Papalexakis, and
  C.~Faloutsos, ``Tensor decomposition for signal processing and machine
  learning,'' \emph{IEEE Transactions on Signal Processing}, vol.~65, no.~13,
  pp. 3551--3582, July 2017.

\bibitem{BCDH+14}
\BIBentryALTinterwordspacing
G.~Ballard, E.~Carson, J.~Demmel, M.~Hoemmen, N.~Knight, and O.~Schwartz,
  ``Communication lower bounds and optimal algorithms for numerical linear
  algebra,'' \emph{Acta Numerica}, vol.~23, pp. 1--155, May 2014. [Online].
  Available: \url{http://journals.cambridge.org/article_S0962492914000038}
\BIBentrySTDinterwordspacing

\bibitem{HK81}
J.~W. Hong and H.~T. Kung, ``{I/O} complexity: The red-blue pebble game,'' in
  \emph{Proceedings of the Thirteenth Annual ACM Symposium on Theory of
  Computing}, ser. STOC '81.\hskip 1em plus 0.5em minus 0.4em\relax ACM, 1981,
  pp. 326--333.

\bibitem{TRG05}
\BIBentryALTinterwordspacing
R.~Thakur, R.~Rabenseifner, and W.~Gropp, ``Optimization of collective
  communication operations in {MPICH},'' \emph{International Journal of High
  Performance Computing Applications}, vol.~19, no.~1, pp. 49--66, 2005.
  [Online]. Available: \url{http://hpc.sagepub.com/content/19/1/49.abstract}
\BIBentrySTDinterwordspacing

\bibitem{ITT04}
D.~Irony, S.~Toledo, and A.~Tiskin, ``Communication lower bounds for
  distributed-memory matrix multiplication,'' \emph{{J. Parallel Distrib.
  Comput.}}, vol.~64, no.~9, pp. 1017--1026, 2004.

\bibitem{SvdG17}
\BIBentryALTinterwordspacing
T.~M. Smith and R.~A. van~de Geijn, ``Pushing the bounds for matrix-matrix
  multiplication,'' arXiv, Tech. Rep. 1702.02017, 2017. [Online]. Available:
  \url{http://arxiv.org/abs/1702.02017}
\BIBentrySTDinterwordspacing

\bibitem{BDHLS12-SS}
\BIBentryALTinterwordspacing
G.~Ballard, J.~Demmel, O.~Holtz, B.~Lipshitz, and O.~Schwartz, ``Brief
  announcement: strong scaling of matrix multiplication algorithms and
  memory-independent communication lower bounds,'' in \emph{{Proceedings of the
  24th ACM Symposium on Parallelism in Algorithms and Architectures}}, ser.
  SPAA '12.\hskip 1em plus 0.5em minus 0.4em\relax New York, NY, USA: ACM, June
  2012, pp. 77--79. [Online]. Available:
  \url{http://doi.acm.org/10.1145/2312005.2312021}
\BIBentrySTDinterwordspacing

\bibitem{DE+13}
\BIBentryALTinterwordspacing
J.~Demmel, D.~Eliahu, A.~Fox, S.~Kamil, B.~Lipshitz, O.~Schwartz, and
  O.~Spillinger, ``Communication-optimal parallel recursive rectangular matrix
  multiplication,'' in \emph{Proceedings of the 27th IEEE International
  Symposium on Parallel and Distributed Processing}, ser. IPDPS '13, 2013, pp.
  261--272. [Online]. Available: \url{http://dx.doi.org/10.1109/IPDPS.2013.80}
\BIBentrySTDinterwordspacing

\bibitem{CDKSY13}
\BIBentryALTinterwordspacing
M.~Christ, J.~Demmel, N.~Knight, T.~Scanlon, and K.~Yelick, ``Communication
  lower bounds and optimal algorithms for programs that reference arrays - part
  1,'' EECS Department, University of California, Berkeley, Tech. Rep.
  UCB/EECS-2013-61, May 2013. [Online]. Available:
  \url{http://www.eecs.berkeley.edu/Pubs/TechRpts/2013/EECS-2013-61.html}
\BIBentrySTDinterwordspacing

\bibitem{BK07}
B.~W. Bader and T.~G. Kolda, ``Efficient {MATLAB} computations with sparse and
  factored tensors,'' \emph{SIAM Journal on Scientific Computing}, vol.~30,
  no.~1, pp. 205--231, December 2007.

\bibitem{PTC13}
A.-H. Phan, P.~Tichavsky, and A.~Cichocki, ``Fast alternating {LS} algorithms
  for high order {CANDECOMP/PARAFAC} tensor factorizations,'' \emph{Signal
  Processing, IEEE Transactions on}, vol.~61, no.~19, pp. 4834--4846, Oct 2013.

\bibitem{CV14}
\BIBentryALTinterwordspacing
J.~H. Choi and S.~V.~N. Vishwanathan, ``{DFacTo}: Distributed factorization of
  tensors,'' in \emph{Proceedings of the 27th International Conference on
  Neural Information Processing Systems}, ser. NIPS '14.\hskip 1em plus 0.5em
  minus 0.4em\relax Cambridge, MA, USA: MIT Press, 2014, pp. 1296--1304.
  [Online]. Available: \url{http://dl.acm.org/citation.cfm?id=2968826.2968971}
\BIBentrySTDinterwordspacing

\bibitem{KU15}
\BIBentryALTinterwordspacing
O.~Kaya and B.~U\c{c}ar, ``Scalable sparse tensor decompositions in distributed
  memory systems,'' in \emph{Proceedings of the International Conference for
  High Performance Computing, Networking, Storage and Analysis}, ser. SC
  '15.\hskip 1em plus 0.5em minus 0.4em\relax New York, NY, USA: ACM, 2015, pp.
  77:1--77:11. [Online]. Available:
  \url{http://doi.acm.org/10.1145/2807591.2807624}
\BIBentrySTDinterwordspacing

\bibitem{SK16}
S.~Smith and G.~Karypis, ``A medium-grained algorithm for distributed sparse
  tensor factorization,'' in \emph{IEEE 30th International Parallel and
  Distributed Processing Symposium}, May 2016, pp. 902--911.

\bibitem{LK+17}
A.~P. Liavas, G.~Kostoulas, G.~Lourakis, K.~Huang, and N.~D. Sidiropoulos,
  ``Nesterov-based parallel algorithm for large-scale nonnegative tensor
  factorization,'' in \emph{IEEE International Conference on Acoustics, Speech
  and Signal Processing (ICASSP)}, March 2017, pp. 5895--5899.

\bibitem{AY16}
\BIBentryALTinterwordspacing
K.~S. Aggour and B.~Yener, ``A parallel {PARAFAC} implementation \& scalability
  testing for large-scale dense tensor decomposition,'' Rensselaer Polytechnic
  Institute, Tech. Rep. 16-02, 2016. [Online]. Available:
  \url{https://www.cs.rpi.edu/research/pdf/16-02.pdf}
\BIBentrySTDinterwordspacing

\bibitem{LW49}
L.~H. Loomis and H.~Whitney, ``An inequality related to the isoperimetric
  inequality,'' \emph{Bulletin of the {AMS}}, vol.~55, pp. 961--962, 1949.

\bibitem{BCCT10}
J.~Bennett, A.~Carbery, M.~Christ, and T.~Tao, ``Finite bounds for
  {H}\"older-{B}rascamp-{L}ieb multilinear inequalities,'' \emph{Mathematical
  Research Letters}, vol.~17, no.~4, pp. 647--666, 2010.

\bibitem{CH+07}
\BIBentryALTinterwordspacing
E.~Chan, M.~Heimlich, A.~Purkayastha, and R.~van~de Geijn, ``Collective
  communication: theory, practice, and experience,'' \emph{Concurrency and
  Computation: Practice and Experience}, vol.~19, no.~13, pp. 1749--1783, 2007.
  [Online]. Available: \url{http://dx.doi.org/10.1002/cpe.1206}
\BIBentrySTDinterwordspacing

\bibitem{ABK16}
\BIBentryALTinterwordspacing
W.~Austin, G.~Ballard, and T.~G. Kolda, ``Parallel tensor compression for
  large-scale scientific data,'' in \emph{Proceedings of the 30th IEEE
  International Parallel and Distributed Processing Symposium}, May 2016, pp.
  912--922. [Online]. Available:
  \url{https://www.computer.org/csdl/proceedings/ipdps/2016/2140/00/2140a912-abs.html}
\BIBentrySTDinterwordspacing

\bibitem{BDKS16}
\BIBentryALTinterwordspacing
G.~Ballard, A.~Druinsky, N.~Knight, and O.~Schwartz, ``Hypergraph partitioning
  for sparse matrix-matrix multiplication,'' \emph{ACM Transactions on Parallel
  Computing}, vol.~3, no.~3, pp. 18:1--18:34, Dec. 2016. [Online]. Available:
  \url{http://doi.acm.org/10.1145/3015144}
\BIBentrySTDinterwordspacing

\end{thebibliography}


\end{document}